\documentclass[fleqn,usenatbib]{mnras}

\usepackage[T1]{fontenc}

\DeclareRobustCommand{\VAN}[3]{#2}
\let\VANthebibliography\thebibliography
\def\thebibliography{\DeclareRobustCommand{\VAN}[3]{##3}\VANthebibliography}

\usepackage{amsmath}
\usepackage{amssymb}
\usepackage{graphicx}
\usepackage{float}
\usepackage{parskip}
\usepackage{caption}
\usepackage{subcaption}
\usepackage{natbib}
\usepackage{newtxtext,newtxmath}
\usepackage{lineno}

\title[A Biotic Habitable Zone: Impacts of Adaptation in Biotic Temperature Regulation]{A Biotic Habitable Zone: Impacts of Adaptation in Biotic Temperature Regulation}

\author[A. E. Nicholson and N. J. Mayne]{
A. E. Nicholson$^{1}$\thanks{E-mail: arwen.e.nicholson@gmail.com}
and N. J. Mayne,$^{1}$
\\
$^{1}$Department of Physics and Astronomy, Faculty of Environment Science and Economy, University of Exeter, EX4 4QL, UK.\\
}

\date{Accepted XXX. Received YYY; in original form ZZZ}

\pubyear{2022}

\begin{document}
\label{firstpage}
\pagerange{\pageref{firstpage}--\pageref{lastpage}}
\maketitle

\begin{abstract}
The search for biosignatures necessitates developing our understanding of life under different conditions. If life can influence the climate evolution of its planet then understanding the behaviour of life-climate feedbacks under extreme conditions is key to determine the `edges' of the habitable zone. Additionally understanding the behaviour of a temperature limited biosphere will help towards formulating biosignature predictions for alien life living under conditions very different to those on Earth. Towards this aim, we extend the `ExoGaia Model' - an abstract model of microbial life living on a highly simplified 0-dimensional planet. Via their metabolisms, microbes influence the atmospheric composition and therefore the temperature of the planet and emergent feedback loops allow microbes to regulate their climate and maintain long term habitability. Here, we adapt the ExoGaia model to include temperature adaptation of the microbes by allowing different species to have different temperature `preferences'. We find that rather than adapting towards the planet’s abiotic conditions the biosphere tends to more strongly influence the climate of its planet, suggesting that the surface temperature of an inhabited planet might be significantly different from that predicted using abiotic models. We find that the success rate for microbial establishment on planets is improved when adaptation is allowed. However, planetary abiotic context is important for determining whether overall survival prospects for life will be improved or degraded. These results indicate the necessity to develop an understanding of life living under different limiting regimes to form predictions for the boundaries of the habitable zone. 
\newline
\end{abstract}




\section{Introduction}

The James Webb Space Telescope (JWST) is already providing an unprecedented window into the chemistry of distant atmospheres \citep{feinsteinERS2022,ahrerERS2022,rustamkulovERS2022,aldersonERS2022}, including detecting products of photochemistry \citep{tsaiERS2022}. JWST is also likely to produce the first detections of the atmospheres of rocky planets in the very near future. The Extremely Large Telescope (under construction), combined with future missions such as the Large Ultraviolet Optical Infrared Surveyor, will extend these capabilities  meaning that searching for biosignatures will become a realistic goal in the coming decades \citep{Snellen:2021, Quanz:2021}. However, a huge amount of progress will be required in our theoretical understanding to both effectively target these resource heavy endeavours and interpret the observations themselves. Searching for signs of life on planets beyond our solar systems requires not only a detailed understanding of the formation of terrestrial planets and the abiotic processes occurring on them, but also a theory for how life interacts with and shapes its planet. Gaia theory \citep{lovelock1965physical, lovelock1974atmospheric, lovelock1990hands} posits that the abiotic components of the planet combined with the biosphere form a complex-system with emergent self-regulatory properties that help explain how Earth has remained habitable and inhabited over geological timescales. Effectively, the Earth and its biosphere form a coupled system.

Multiple models have been developed to explore how Gaian regulation can occur in a life-environment coupled system and have been used to study the response of such systems to perturbations, both internal (e.g. via biological mutations) and external (e.g. solar luminosity changing over time) \citep{watson1983biological,downing1999simulated, williams2007flask, wood2008daisyworld, worden2010notes, nicholson2017multiple, gaianhabzone}. Life on Earth has played a primary role in shaping the chemistry of the atmosphere and oceans \citep{lenton2013revolutions}, and evidence indicates it might even play a role in the continental coverage of Earth \citep{honing2014biotic, honing2016continental} as well as the formation and maintenance of plate tectonics, which are vital for climate regulation \citep{lenardic2016solar, plank2019subducting}. Models indicate that without life the Earth would be significantly warmer than it is today (up to $45^{o}C$ warmer) \citep{schwartzman1989biotic}, as life hugely accelerates the removal of $CO_{2}$ from the atmosphere via chemical weather thus reducing surface temperatures. Additionally, accumulation of oxygen in the atmosphere of the early Earth due to life is thought to have triggered planet wide glaciation events during the Proterozoic (between 2.5 billion to 541 million years ago) when the Earth would have been covered from poles to equator by ice for millions of years at a time \citep{kopp2005paleoproterozoic}.

As life has played such a vital role in the climate evolution of our own planet, Gaia Theory will very likely have applications in the search for `biosignatures' - remotely detectable signs of alien life. At the very least the story of our own planet suggests that biotic processes can not be neglected when considering the evolution and, crucially, atmospheric composition of life-hosting planet. The search for biosignatures tends to focus on finding chemical signatures in the atmosphere of a planet that indicate activity by life \citep{sousa_silva_2018}. Obtaining observations of these signatures is a resource intensive process, therefore, a framework is required to effectively guide the candidate selection \citep{seager2014}. The habitable zone defines the radii about a star where a planet is theoretically capable of hosting surface liquid water and therefore of potentially hosting life as we know it \citep{kasting1993habitable, kopparapu2013revised, abe2011habitable}. The vast majority of work defining the habitable zone focuses solely on abiotic processes \citep[e.g.,][]{paradise2017}, and such studies have demonstrated the dependence of the habitable zone on many factors including the age and class of the host star \cite{ramirez2016habitable}, planetary mass \cite{kopparapu2014habitable}, planetary atmospheric composition \cite{pierrehumbert2011hydrogen}, and the surface water content of the planet \cite{abe2011habitable}. Attempts have also been made to define an abiogenesis zone \citep{rimmer2018}, within which the early building blocks of life might be generated. However models indicate that planetary formation might not be deterministic \citep{lenardic2016solar, weller2018evolution, lenardic2019toward}, and if life can strongly impact the climate of its planet, it could alter the boundaries of the habitable zone to beyond those calculated using abiotic models alone. Gaia Theory would predict that the Habitable Zone is a property of not only the host star and bulk planet characteristics but also of the life that could emerge on a planet \citep{chopra2016case} and so suggests that the boundaries of a `biotic' habitable zone might differ significantly to those of an abiotic one. Additionally the habitable zone for different types of planets, for example $H_{2}$ dominated super Earths, could differ significantly compared to the habitable zone for Earth-like planets \citep{pierrehumbert2011hydrogen,seager2013exoplanet}. 

As our sample size of potentially life-bearing planets increases, we must begin to incorporate understanding of life--climate interactions in our definition of any habitable zone. Moreover, a more developed theory of life--climate interaction will enable us to begin to determine how the probability of a planet being inhabited varies across the habitable zone. This aim is, no doubt, hugely complex, but we must begin to take the first steps, without which, progress towards this goal is impossible.

Models of Gaian regulation where the life-environment coupled system can switch between limiting regimes (e.g. by limited by nutrient availability such as $H_{2}$ availability or by having their metabolism limited by an environmental parameter such as temperature), demonstrate different system behaviours for different limiting regimes \cite{gaianhabzone, nicholson2017multiple, wood2008daisyworld}. This could indicate that climate regulation on planets at the edges of the habitable zone might function differently than on those towards its centre. A temperature limited biosphere has been proposed as a mechanism to produce a stabilising feedback on the climate of potentially habitable $H_{2}$ dominated greenhouse planets, thus significantly extending the habitable zone for these planets \citep{abbot2015proposal}. Previous work \citep{nicholson2022predicting} has taken steps to form predictions for biosignatures for nutrient-limited biospheres, and found that the ability of the biosphere to exploit its limiting resource was more fundamental to the observable biosignature than the details of any potential alien biology. A biomass based model of determining biosignature plausibility has also been proposed by \cite{seager2013biomass}. Similar work will be required to form predictions for biosignatures of temperature limited biospheres. In these cases potential biosignatures would directly depend on the temperature at which the planet's climate stabilises. Therefore not only are such scenarios of interest when forming predictions for potential biosignatures for biospheres under different limiting conditions, but are also of interest for attempting to determine the edges of the habitable zones. Multiple biosignature hypotheses, in addition to accurate abiotic models will be needed to compare to observational data to begin the process of identifying distant signs of alien life. 

Towards these aims, this work builds on the ExoGaia model \citep{nicholson2018gaian} - an abstract model where a temperature-sensitive microbial biosphere is introduced to a highly simplified planet. In this model, microbes are able to `catch' a window of habitability and maintain habitable temperatures for long time periods by regulating the planet's atmospheric composition and hence the climate. In the absence of life, most ExoGaia planets would quickly revert to inhospitable temperatures. We extend ExoGaia to explore how including microbe adaptability to climatic conditions (a feature previously neglected in the model) impacts the emergent regulatory behaviour, and in turn the planetary surface temperature over time. Different life forms on Earth are adapted to a wide range of environments with varying pH values, temperatures, salinity etc., and allowing varying temperature preferences between model lifeforms has been found to be important for certain regulation mechanisms to emerge in other Gaian models, e.g. \cite{dyke2010daisystat}. Therefore including this feature is important to fully understand the behaviour of the emergent regulation in the ExoGaia model.

The rest of this paper is laid out as follows: Section \ref{sec:methods} sets out model design, Section \ref{Section:experimentsetup} explains the experiment setup, Section \ref{Section:hab_regulation_original} details the emergent temperature regulation in the ExoGaia model, Section \ref{Results} presents the results from the experiments, and  Section \ref{discussion} discusses the results in detail. Future work is discussed in Section \ref{Section:further_work}.

\section{Model setup}
\label{sec:methods}

Here we will detail the ExoGaia model as used in this work, however a full and more complete description of the model where microbe temperature preferences are kept constant, can be found in \citet{nicholson2018gaian}. The model comprises of a planet represented by a zero dimensional atmosphere composed of gaseous chemicals, and this planet orbits a star that provides an influx of radiation. The model life consists of simple microbes that live on an implied surface on the planet.

\begin{figure}
\centering
\includegraphics[width=0.45\textwidth]{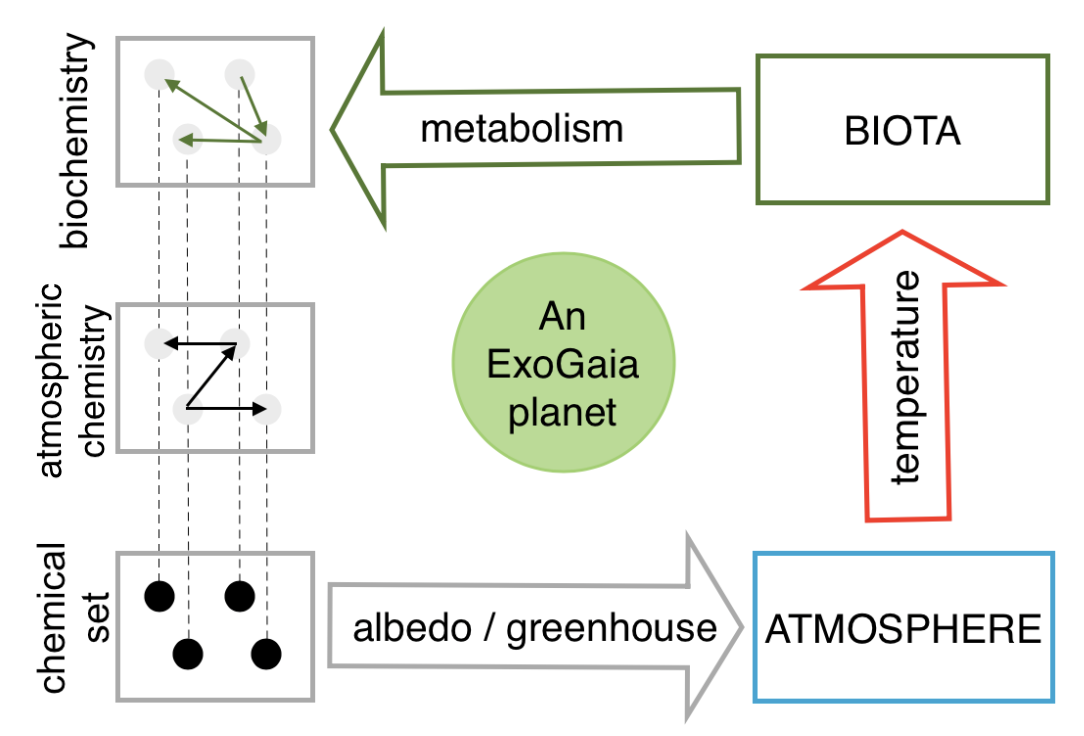}
\caption{Schematic of an ExoGaia planet}
\label{fig:schematic}
\end{figure}

A schematic of an ExoGaia planet is shown in Figure \ref{fig:schematic}. An ExoGaia planet consists of a zero-dimensional atmosphere, composed of abstract representations of chemical species, with an implied surface which life can live on. These planets orbit a star that provides incoming radiation that warms the planet. The chemical composition of the atmosphere determines the overall albedo or greenhouse effect of the atmosphere and thus impacts the planetary temperature. 

Background atmospheric chemical reactions represent abiotic processes that convert one chemical to another. Life then adds biochemistry by converting one chemical to another via their metabolism. By changing the composition of the atmosphere, both the atmospheric chemistry and the biochemistry influence the temperature of the ExoGaia planet. Model microbes have temperature dependant metabolisms, where their ability to consume their food source is at a maximum when the planet's surface temperature is equal to their `preferred temperature', and their ability to consume their food source decreases smoothly and symmetrically as the temperature moves away from this value. This is an idealised way of capturing the behaviour of real lifeforms' metabolisms, the rate of which are impacted by the temperature of their environment (e.g. \cite{price2004temperature}).

\subsection{Atmospheric chemistry}
\label{exogaia_planets}

We simulate planets with eight possible atmospheric chemicals and set two to have an `inflow' in an abstract representation of outgassing from volcanoes, in line with previous work \citep{nicholson2018gaian, alcabes2020robustness}. Chemicals that do not have an inflow can only materialise in the atmosphere if abiotic chemical reactions or microbe metabolisms produce them. All chemicals experience a constant rate of out-flux from the atmosphere. For simplicity atmospheric chemical reactions are limited to have only one reactant and one product e.g. $A \rightarrow B$ but not $A + B \rightarrow C$, and atmospheric chemical reactions are uni-directional e.g. if A $\rightarrow$ B then B $\nrightarrow$ A.

Each atmospheric chemical has a randomly assigned albedo or greenhouse effect on the atmosphere, and the impact of a chemical on the planetary temperature scales with its abundance, see \cite{nicholson2018gaian} for further details. Therefore the chemical composition of the atmosphere, combined with the incoming radiation of the planet's star, determine the planetary temperature. This setup captures the basic elements of a planetary climate. 

For simplicity, and in keeping with previous work \citep{nicholson2018gaian, alcabes2020robustness}, we keep the atmospheric chemistry to be temperature independent so that the rate of each chemical reaction does not change when the planetary temperature changes. Although real world geochemistry is temperature sensitive, by excluding this we can be confident that any emergent regulation in the  model is due to the temperature sensitivity of the microbial life and not due to temperature sensitive abiotic chemistry.

\subsection{Microbial life}
\label{microbial_life}

Model microbes consume and excrete atmospheric chemicals to obtain energy for their metabolic processes and biomass building. Microbe metabolisms are kept simple so that each species consumes only one food source $S$ and excretes one waste product $W$ such that $S \neq W$. Different species of microbe have different metabolisms.

Microbes convert consumed food into biomass which they require for their maintenance cost (i.e. homeostasis), and must accumulate in order to reproduce. The conversion from food to biomass is an inefficient process with the proportion of food not converted to biomass instead being excreted as waste product. If a microbe's biomass drops below a threshold, it will die via starvation. There is also a constant rate of `random' death representing death via means other than starvation. This is set to $p_{kill} = 0.005$ and kept constant throughout this work.

The ability of microbes to consume their food source is temperature dependant, therefore the microbes add temperature dependant biochemical reactions to the atmospheric chemistry. Microbes have a temperature $T_{pref}$ at which their growth rate is at a maximum. As the temperature of the planet $T_{planet}$ moves away from $T_{pref}$ the microbes growth rate drops and where the distance between $T_{pref}$ and $T_{planet}$ becomes too large, microbes are unable to consume their food source. In this scenario unless the environment improves, the microbe's metabolisms will halt and they will be unable to generate the biomass they require for cell maintenance and reproduction and risk starving to death. Microbes face an additional constant background probability of death representing microbes dying from factors other than starvation. As before, although abstract and highly idealised, this setup does capture the primary behaviours of simple lifeforms. Once microbes die we assume that their bodies are lost (e.g. sink to the bottom of the ocean) and the contained biomass is lost from the system. We keep the ecosystem simple and neglect to include secondary consumers, therefore ExoGaia is best thought of as a representation of early life emerging on a young planet before more complex ecological systems have evolved. Including secondary consumers could impact the emergent regulation in ExoGaia and we leave exploring this to future work.

\section{Experiment setup}
\label{Section:experimentsetup}

The experiment setup we adopt is such that in the absence of an atmosphere, the planet is too cold for life. Initially we set the planet up so that there are no chemicals in the atmosphere. At the start of the experiment, chemicals that have an abiotic inflow source begin to build up in the atmosphere at a constant rate. The abiotic atmospheric chemistry then converts some of these `inflow' chemicals into others, also at a constant rate. As the composition of the atmosphere changes the temperature of the planet changes.

Time in the ExoGaia model is recorded in abstract units of `timesteps'. A timestep is determined by the total microbe population, and the number of different biological processes that microbes can undergo. Therefore within a timestep the abiotic processes are only implemented once whereas the biological processes will be implemented a number of times as determined by the total population. This is to capture that microbes would all be interacting with their environment in parallel not sequentially. This approach has been used in previous abstract models of life-environment interaction (e.g. \cite{becker2014evolution, williams2007flask}). As we are studying the long-term stability of abstract life on a given planet we effectively evolve the model to a `steady state' condition and the timestep lengths do not correspond to a real world value. A single timestep in the ExoGaia consists of the following computational steps:


\begin{enumerate} 
\item Update atmospheric composition via inflows, outflows, and atmospheric chemistry.
\item Update planet temperature given new atmospheric composition.
\item If the planet is inhabited, update the microbe population.\\ For a total microbe population of $N$ (at the start of the timestep), randomly select a microbe $N$ times and for each microbe perform the following steps:
\begin{itemize}
    \item Apply chance of random death or starvation (if biomass is insufficient).
    \item Expend maintenance cost.
    \item Reproduce if sufficient biomass available. 
    \item Consume food if available.
    \item Excrete waste product. 
\end{itemize}
\end{enumerate}
Then return to the start of the process for the next timestep. 

Temperature in the ExoGaia model is in abstract units, and the microbes' preferred temperature is not meant to reflect a realistic temperature at which biology thrives. As ExoGaia is an abstract model of a planet-life coupled system, it cannot be used to make quantitative predictions for real planets, but instead is designed to investigate the qualitative behaviour of such a system. To reinforce this in the original model we chose $T_{pref} = 1000$ \citep{nicholson2018gaian}.

In the original ExoGaia model \citep{nicholson2018gaian}, $T_{pref}$ was kept constant for all microbe species and species only differed in their metabolisms (i.e. the chemical they consume and excrete). Here we allow $T_{pref}$ to differ between species (within a defined allowed $T_{pref}$ range) to investigate how this impacts the model dynamics. We explore a scenario where the allowed $T_{pref}$ values for all species are defined at the start of the experiment by drawing from a normal distribution centered about $T^{C}_{pref} = 1000$ with variance of $50$. Therefore, the allowed $T_{pref}$ range is symmetric.

We `seed' an ExoGaia planet with a single species of life and for both sets of experiments we set $T_{pref} = 1000$ for this first emergent species. Life is seeded on the ExoGaia planet either when $T = T_{pref} = 1000$, where $T$ is the global temperature of the planet, or after $5 \times 10^{4}$ timesteps, whichever occurs first. The experiment ends $5 \times 10^{5}$ timesteps after life is seeded onto the planet.

In this study we use the same planetary setups as those explored in \cite{nicholson2018gaian}. The microbe parameters (other than allowing for $T_{pref}$ to vary between species) are also identical to that work. Therefore in this paper, the simulations where $T_{pref}$ is kept constant for all species are identical to the results presented in \cite{nicholson2018gaian}. We then use these experiments as a baseline to compare with experiments allowing $T_{pref}$ to vary to investigate how this change impacts the life-environment coupled systems. 

\section{Temperature regulation}
\label{Section:hab_regulation_original}

Regulation of the planetary temperature emerged in the original ExoGaia model due to the feedbacks between temperature and microbe metabolism. This regulation emerges and is maintained via the combined result of the abiotic chemical reactions and the metabolisms of microbes rapidly recycling chemicals and stabilising the climate. As the ExoGaia model is zero-dimensional, the entire biosphere experiences the same environment. The experiment setup of the model is such that in the absence of an atmosphere, the planet is too cold for life. As atmospheric chemicals build up in the atmosphere, the surface temperature changes. If the atmosphere is overall insulating, which is the default ExoGaia set-up, then the temperature will rise \citep[versions of the model where the atmosphere is overall cooling are explored in ][]{nicholson2018gaian}. The habitable range of temperatures is much lower than the abiotic steady-state temperature for the vast majority of the ExoGaia planet setups explored. This is chosen to be consistent with the prediction that the Earth would be warmer in the absence of life \citep{schwartzman1989biotic} and would be much colder without its insulating atmosphere, this setup serves as a very simplistic and abstract representation of an Earth-like planet.

When life is introduced to an ExoGaia planet it changes the atmospheric chemistry and thus the temperature. An individual microbe's metabolism might be `cooling' or `warming' depending on which chemical it consumes and which it excretes, however the end result of this metabolism will depend on whether the chemical excreted is further transformed via other metabolisms or abiotic atmospheric chemical reactions. On the vast majority of ExoGaia planetary setups, the temperature can only stabilise at a habitable level once complete recycling of the atmospheric chemistry is established, i.e. no chemicals are building up in an uncontrolled manner. This recycling will emerge as the combination of the biosphere's metabolisms together with the abiotic chemistry. Once complete recycling is established the biosphere influences the entire atmosphere even if a metabolism doesn't directly interact with a particular atmospheric chemical. As the habitable temperature range is much lower than the abiotic steady state temperature for most planets, the biosphere must `catch' a window of habitability and maintain it to survive. On a planet with an overall insulating atmosphere, when a biosphere has established atmospheric regulation, by reducing the abundance of atmospheric chemicals in the atmosphere the biosphere will have an overall cooling effect on its planet.

\begin{figure}
\centering
\includegraphics[width=0.45\textwidth]{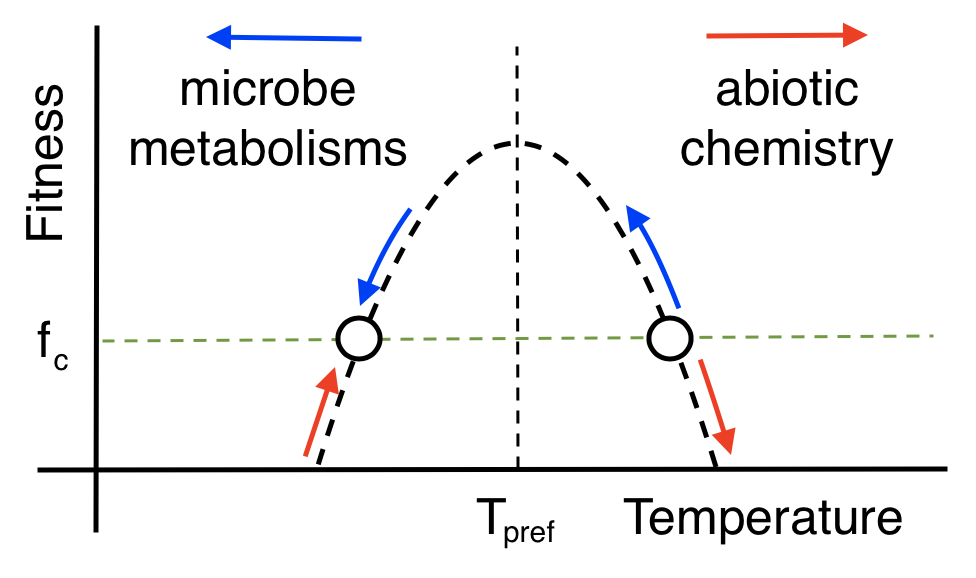}
\caption{Schematic of the fitness, $f_{c}$, of a microbe metabolism showing a symmetric curve around an optimum temperature of $T_{pref}$. In our model setup the atmosphere tends to warm the planet, in the absence of life, with the biosphere acting to cool the planet, as is the case for Earth.}
\label{fig:original_regulation}
\end{figure}

The maximum `fitness' of the microbes, i.e. where their growth rate is at a maximum is defined as $T_{pref}$. A stable population is reached where the death rate of the microbes matches the birth rate, and we label this fitness $f_{c}$. The temperature sensitivity of microbes is symmetric about $T_{pref}$ and there are two possible solutions for $f_{c}$, one where $T > T_{pref}$ and one where $T < T_{pref}$ as shown in Figure \ref{fig:original_regulation}. However only one of these solutions is actually stable due to the feedback loops present in the system between microbe fitness and temperature.

\begin{figure}
\centering
\includegraphics[width=0.45\textwidth]{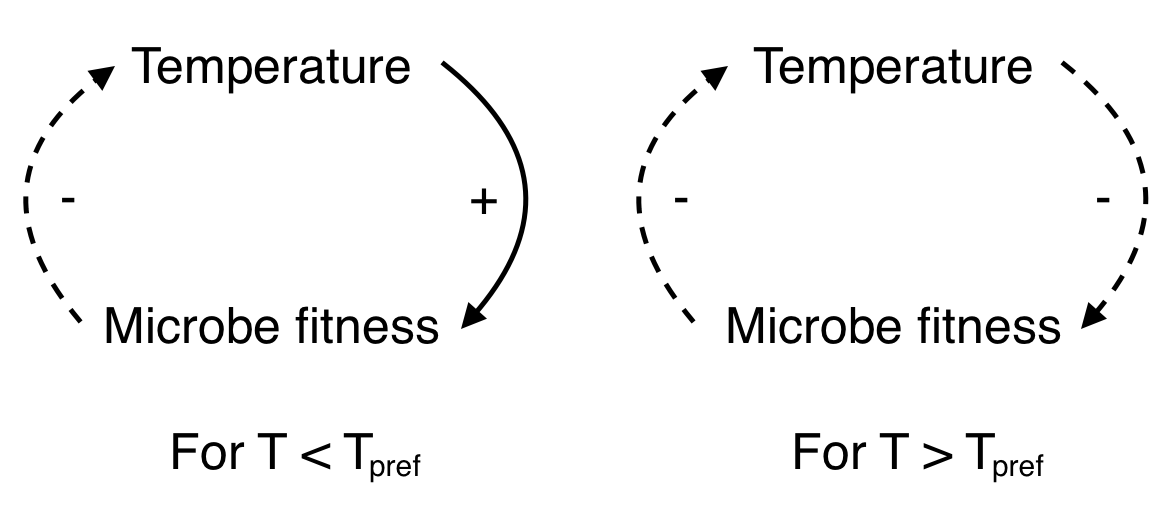}
\caption{Diagrams capturing the feedback loops between life and the planetary temperature for cases hotter and cooler than the preferred temperature for a given microbe.}
\label{fig:feedbacks}
\end{figure}

Figure \ref{fig:feedbacks} shows the feedback loops between temperature and microbe fitness for scenarios where $T < T_{pref}$ and $T > T_{pref}$. A solid arrow indicates that an increase in the source leads to an increase in the sink (e.g. an increase in temperature leads to an increase in microbial fitness), and a dashed arrow indicates that an increase in the source leads to a decrease in the sink. If the overall sign of the feedback loop is positive, then the feedback is amplifying, if the overall sign is negative then the feedback loop is stabilising.

When $T < T_{pref}$ an increase in the temperature will lead to an increase in the microbes fitness as $T$ moves closer to $T_{pref}$, and this will cause an increase in metabolic activity and an increase in population. But as the biosphere collectively acts to cool the planet, by removing atmospheric chemicals via microbe metabolisms, an increase in metabolic rate / total population will lead to a reduction in temperature and so dampen the previous perturbation. If, when $T < T_{pref}$ the temperature reduces, $T$ moves further from $T_{pref}$ and so lowers the microbes' fitness and reduces the population. Therefore the impact of the microbes on the atmosphere is reduced, and the background abiotic chemistry, which overall warms the planet will increase $T$, again dampening the initial perturbation. Therefore when $T < T_{pref}$ the temperature is regulated. 

When $ T > T_{pref}$ these feedback loops function differently. Now an increase in $T$ will reduce the microbes fitness and so reduce their cooling impact on the planet and so further increase $T$ - amplifying the initial perturbation. If however a perturbation reduces $T$, then $T$ is closer to $T_{pref}$ and so the microbes fitness will increase, increasing their population and thus cooling the planet further - again, amplifying the initial perturbation. Therefore the microbial biosphere can only regulate the planet temperature for $T < T_{pref}$ for scenarios where the atmosphere is overall insulating.

As all microbes share a global environment with a single temperature, when $T_{pref}$ is constant for all species every microbe will experience the same fitness regardless of its individual metabolism. Therefore it is the overall impact of the biosphere on its environment that is important for climate regulation and the above feedback loops function on the biosphere as a whole. Therefore if we run identical planet set ups multiple times, the emergent biospheres within each experiment might alter the atmospheric composition differently due to different microbe communities, however regulation will be established at the same temperature. Any biosphere that fails to establish regulation will go extinct. 

Alternative scenarios were explored in \citep{nicholson2018gaian} where instead of insulating, the atmosphere is overall cooling, and the planets orbit a hotter star. In this scenario the biosphere collectively acted to overall warm the planet as in this scenario when atmospheric regulation is in place, reducing the abundance of atmospheric chemicals reduces the albedo of the atmosphere, reflecting less of the sun's radiation away. Under this scenario the feedback loops above work in the same way, but reversed, now temperature regulation can only take place \textit{above} $T_{pref}$ and below $T_{pref}$ a runaway feedback loop can occur where the planet temperature drops to inhospitably cold temperatures.

An individual microbe metabolism might have a warming impact on the planet, e.g. by consuming a cooling atmospheric chemical and excreting an insulating one that is not recycled via abiotic processes. However, in isolation this scenario is short lived as the climate cannot stabilise, and so temperatures will climb and the microbes will go extinct. The climate can only stabilise when recycling of atmospheric chemicals is established by the biosphere. With recycling in place for an overall insulating atmosphere, the overall impact of the biosphere is then to cool by removing atmospheric chemicals, no matter the individual metabolisms of the microbes that form the biosphere \citep{nicholson2018gaian}. 

\begin{figure*}
\centering
\begin{subfigure}{.49\textwidth}
  \centering
  \includegraphics[scale=0.49]{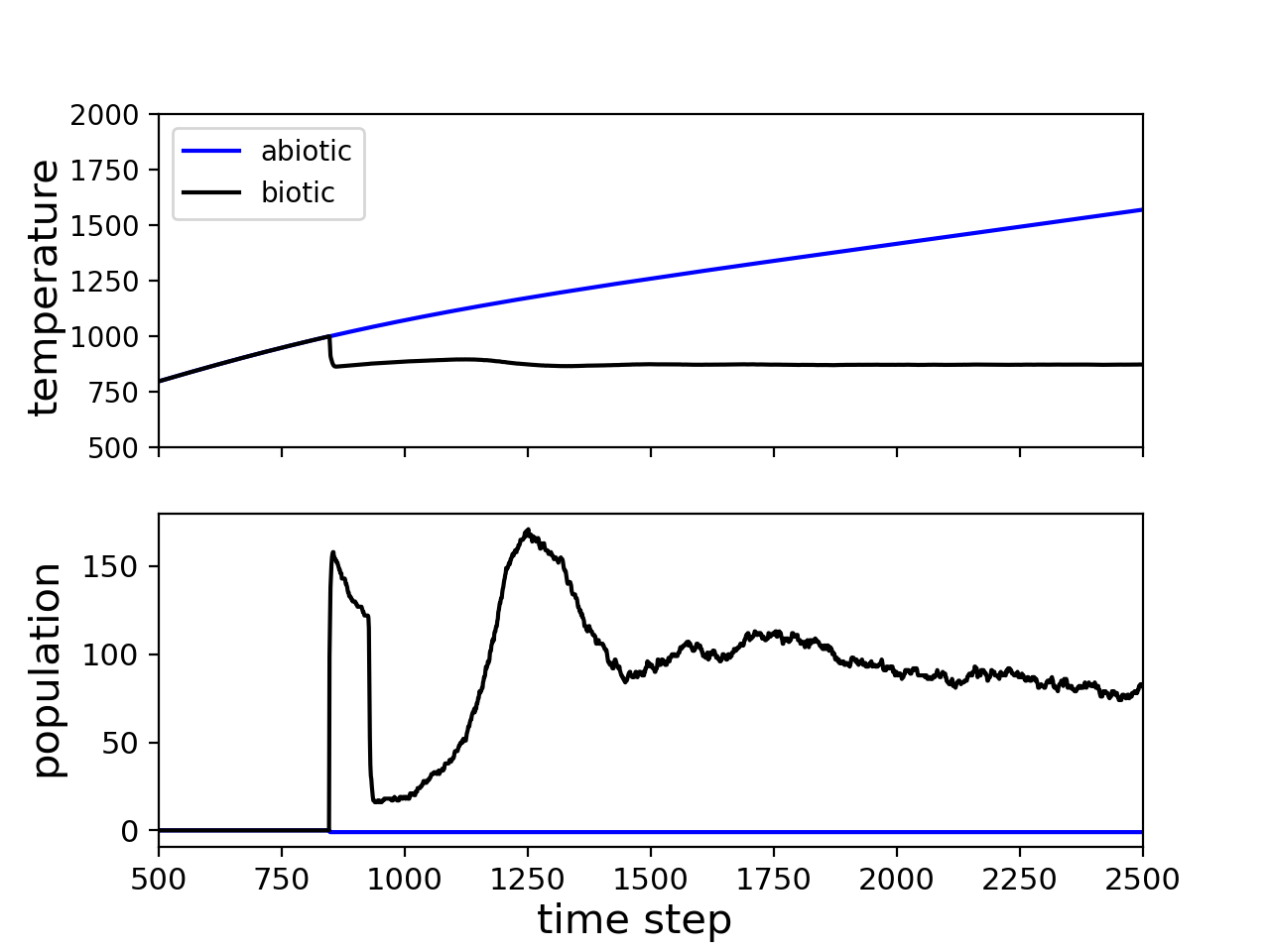}
  \caption{Behaviour soon after seeding}
  \label{fig19:a}
\end{subfigure}%
\begin{subfigure}{.49\textwidth}
  \centering
  \includegraphics[scale=0.49]{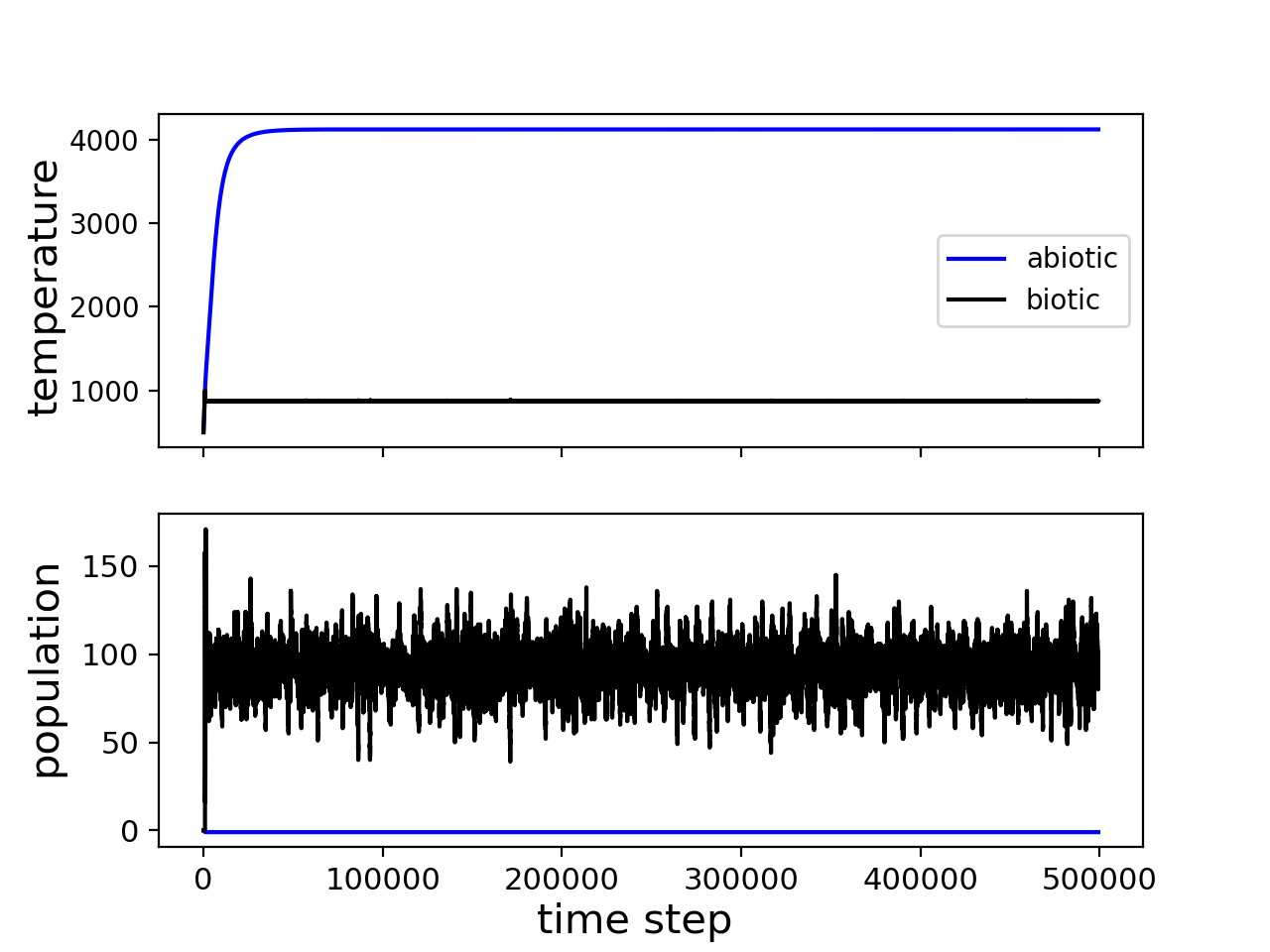}
  \caption{Long term behaviour}
  \label{fig19:b}
\end{subfigure}
\centering
\caption{Two examples using the same planetary set up. One where life is seeded (labeled `biotic') and the other where no life is introduced (labeled `abiotic').}
\label{fig:biotic_vs_abitoic}
\end{figure*}

Figure \ref{fig:biotic_vs_abitoic} shows two experiments on the same planet, one where life is introduced and takes hold on the planet, and one without life. We seed with life once $T = T_{pref}$. We find that without life, the planet quickly increases in temperature, as greenhouse gases build up in the atmosphere, and the planet quickly becomes inhospitable. For the majority of ExoGaia planets explored in the original work \citep{nicholson2018gaian}, the `equilibrium temperature' of the planet - that is the temperature of a purely abiotic planet once the atmosphere has stabilised - is far far higher than $T_{pref}$. With life however, the microbial biosphere is able to `catch' the window of habitability, and by consuming parts of the atmosphere and thus preventing the uncontrolled buildup of greenhouse gases, are able to maintain habitability over long time spans. This regulation emerges due to the feedback loops between microbial metabolic activity and the planet's temperature outlined in Figures \ref{fig:original_regulation} and \ref{fig:feedbacks}. This behaviour links to hypotheses that planetary evolution is not a deterministic process, and that small perturbations early on in a planet's history could dramatically impact its later climate \citep{chopra2016case, lenardic2016solar}.

Work by \cite{alcabes2020robustness} extended the ExoGaia model to introduce external perturbations, both rapid and gradual, by changing the incoming `solar luminosity' to an ExoGaia planet. They found that while introducing perturbations could disrupt regulation on a model planet, overall the regulation of the environment by the biosphere was robust in the face of perturbations. In this work we will investigate how allowing $T_{pref}$ to differ for different microbe species impacts the dynamics of the system.

ExoGaia planets fall into five categories that are defined by the seeded life's ability to successfully establish on their planet, and the ability of the system to support life until the end of the experiment. Establishment success is defined as life surviving for $t = 1000$ time steps after it is seeded. This length of time was chosen as it is long enough for microbes experiencing inhospitable conditions to starve to death as their biomass reserves are depleted. Long term habitability success is defined as life surviving for $t = 500,000$ time steps - the length of each experiment. The planet classifications are thus defined as follows:

\begin{itemize}
    \item Abiding - life always successfully establishes itself on the planet and survives to the end of the experiment.
    \item Bottleneck - if life is successful in establishing itself on the planet then the biosphere survives until the end of the experiment.
    \item Critical - the biosphere is prone to going extinct at any point during the experiment.
    \item Doomed - although the equilibrium temperature of the planet is in the habitable range, life in unable to establish itself the planet.
    \item Extreme - the temperature of the planet never enters the habitable range for life.
\end{itemize}

These planet categories are defined and explored in detail in \cite{nicholson2018gaian}. The `connectivity' of the background atmospheric chemistry, i.e. the number of abiotic chemical reactions, plays a large role in determining the long term habitability prospects of a planet. For ExoGaia planets with a well connected abiotic chemistry, with preexisting recycling loops between atmospheric chemicals, life can more easily gain `control' of the atmosphere and the system is less prone to perturbations. If there are fewer or no abiotic recycling loops between atmospheric chemicals, then life must effectively fill in more of the `blanks' to achieve atmospheric regulation and prevent the uncontrolled build up of any chemical, which would disrupt the habitability of the planet. In these scenarios, as the microbes are temperature sensitive, the `connections' that microbes form, for example recycling chemical A to chemical B, are also temperature sensitive, so if the population of the microbes drops, the recycling of atmospheric chemicals can become disrupted which can lead to a total breakdown of atmospheric regulation and therefore total extinction. 

\begin{figure}
\centering
\includegraphics[width=0.45\textwidth]{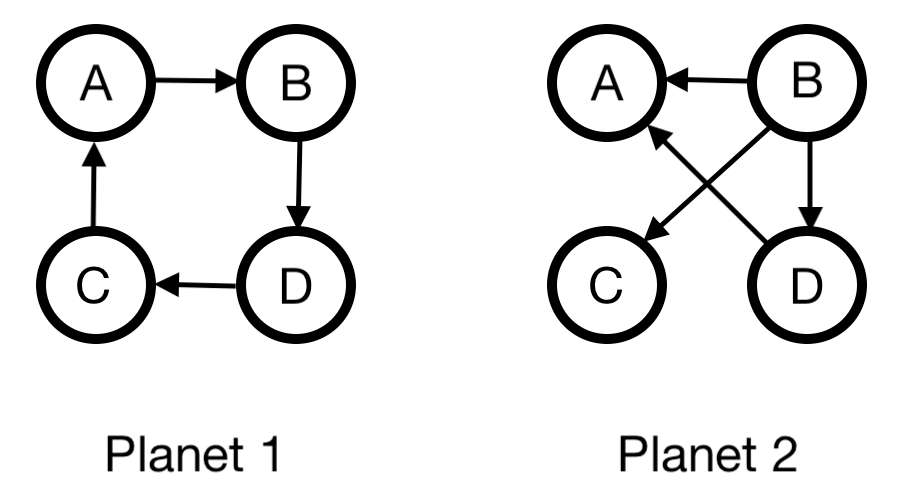}
\caption{Schematics showing two different planets (1 \& 2) with the same atmospheric chemicals (A, B, C, D), but different 'links' transforming chemicals into one another representing atmospheric chemical processes occurring on the planet.}
\label{fig:example_planets}
\end{figure}

Figure \ref{fig:example_planets} shows two example planet abiotic chemistry set ups, with 4 atmospheric chemicals, labeled $A$ to $D$, instead of 8 (as is the case for the simulations in this work) for simplicity. Arrows between chemicals indicate abiotic chemical reactions converting one chemical to another at some set rate. For Planet 1's set up, we see that a feedback loop including all the atmospheric chemicals exists, and so any life emerging on the planet can amplify this preexisting feedback loop to quickly and easily establish atmospheric regulation. For Planet 2 however there are two places where chemicals can build up in the atmosphere without recycling taking place - chemicals $A$ and $C$. Therefore for complete recycling of Planet 2's atmospheric chemicals, more biotic links are required to `close' the gaps in the abiotic network. As biotic links are temperature sensitive they are prone to perturbations and so a recycling network that depends on numerous biotic links to be complete is more prone to fluctuations and thus breaking down. Therefore, as more biotic links are required to regulate Planet 2's atmosphere (and thus temperature) when compared to Planet 1, atmospheric regulation in the case of Planet 2 is more prone to breaking down and the long term habitability prospects of Planet 2 will be lower than that of Planet 1.

\section{Results}
\label{Results}

Here we present the results from our experiments. We compare identical planet simulation set ups and compare cases where $T_{pref}$ is fixed for all microbe species in one experiment, and allowed to vary in the other. We then look at the behaviour of introducing varying $T_{pref}$ over 100 experiments for 100 different planetary set ups to investigate the average behaviour of planets hosting life with varying $T_{pref}$.

\subsection{Individual simulations}
\label{Section:individual_simulations}

\begin{figure*}
\centering
\begin{subfigure}{.49\textwidth}
  \centering
  \includegraphics[scale=0.49]{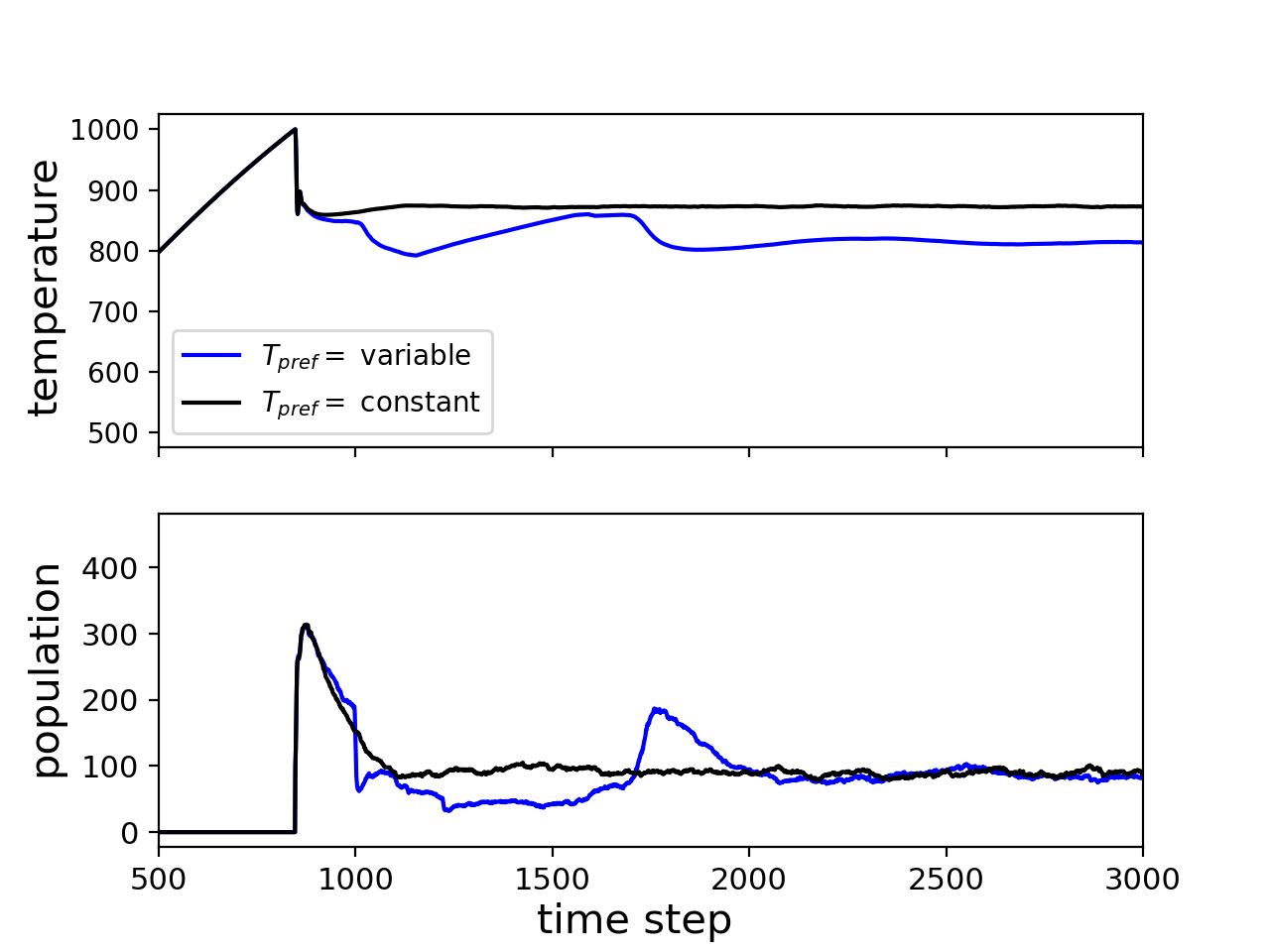}
  \caption{Behaviour soon after seeding Abiding}
  \label{fig:Abiding_example_a}
\end{subfigure}%
\begin{subfigure}{.49\textwidth}
  \centering
  \includegraphics[scale=0.49]{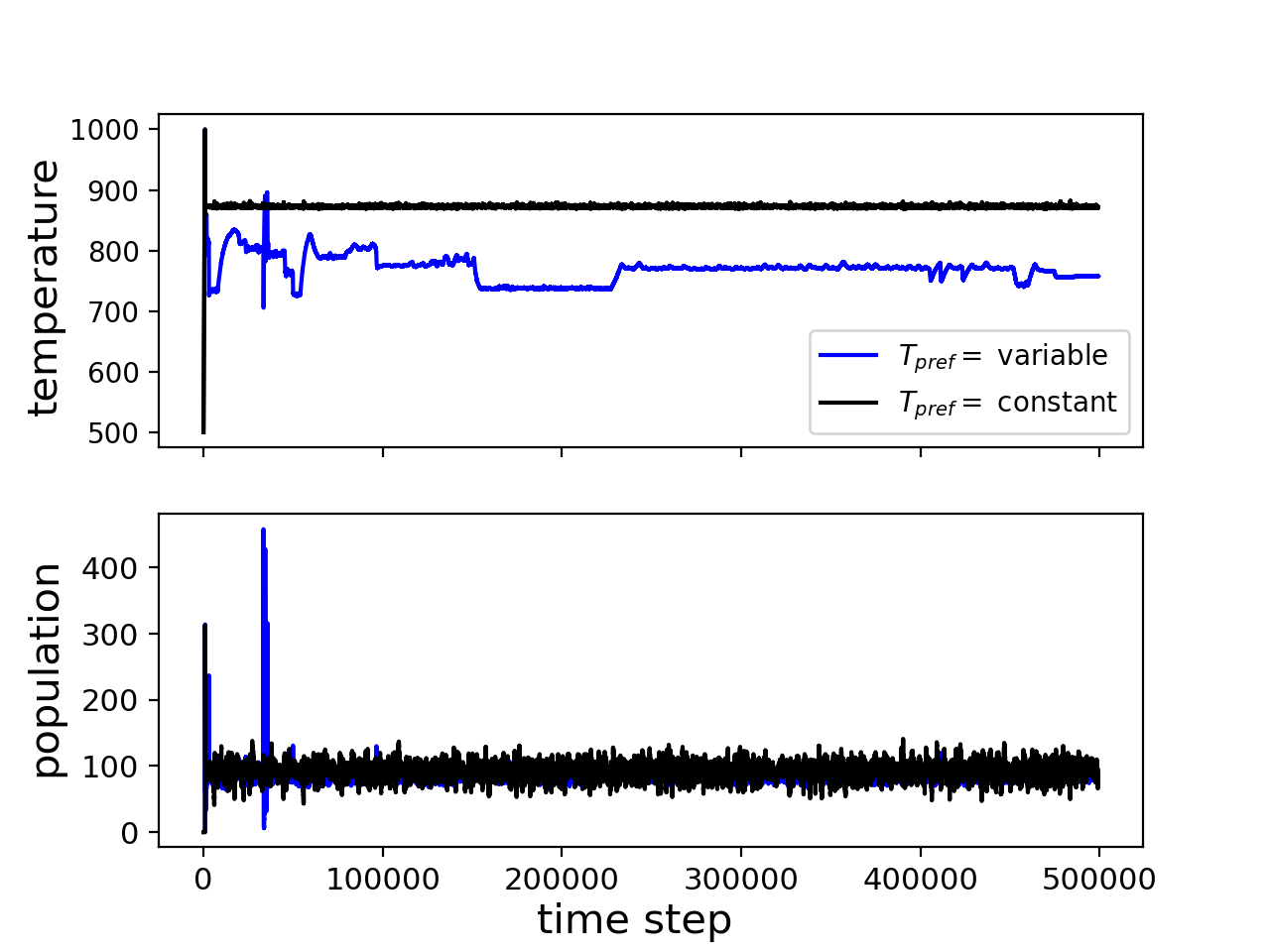}
  \caption{Long term behaviour Abiding}
  \label{fig:Abiding_example_b}
\end{subfigure}

\centering
\begin{subfigure}{.49\textwidth}
  \centering
  \includegraphics[scale=0.49]{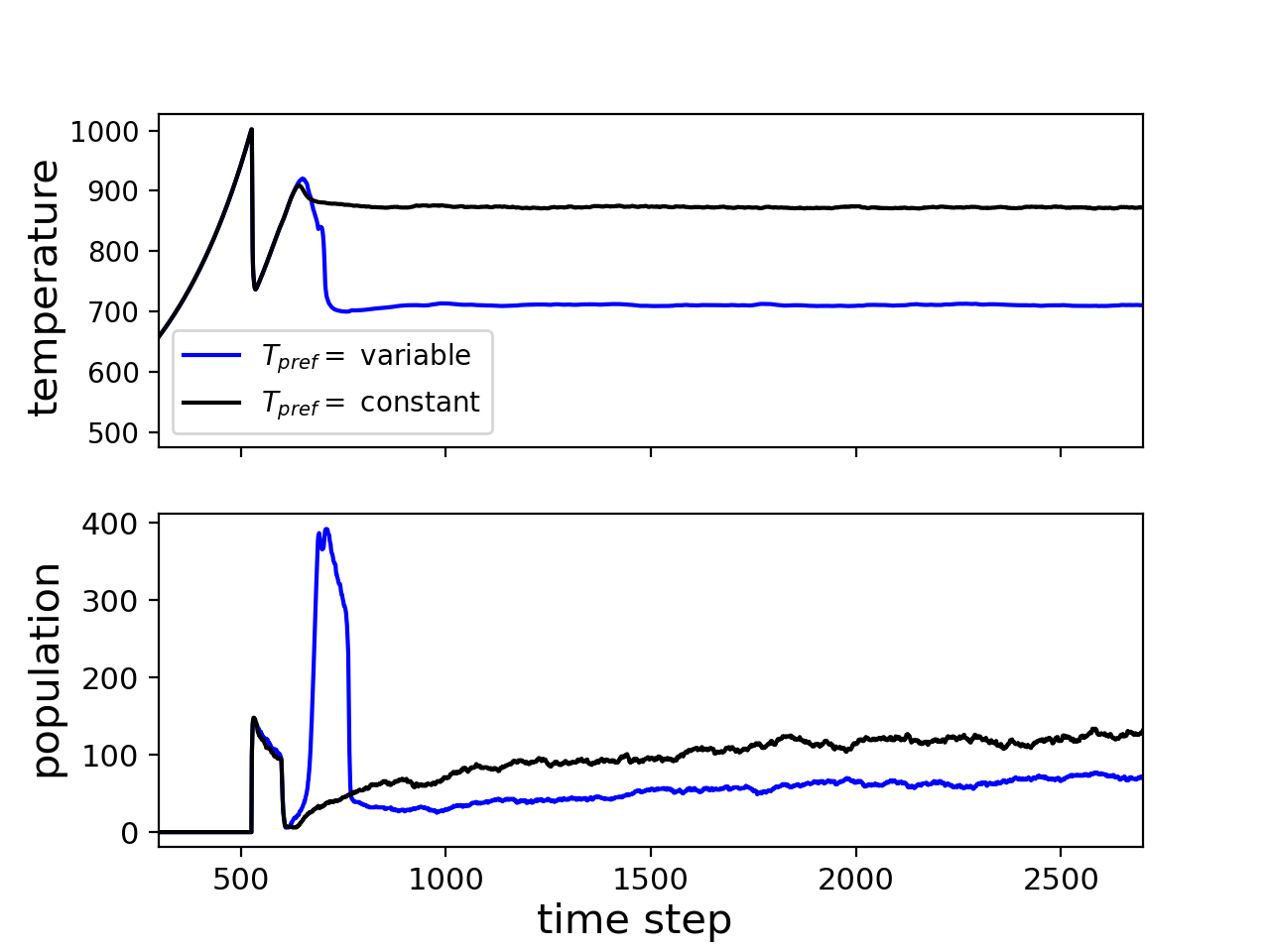}
  \caption{Behaviour soon after seeding Bottleneck}
  \label{fig:Bottleneck_example_a}
\end{subfigure}%
\begin{subfigure}{.49\textwidth}
  \centering
  \includegraphics[scale=0.49]{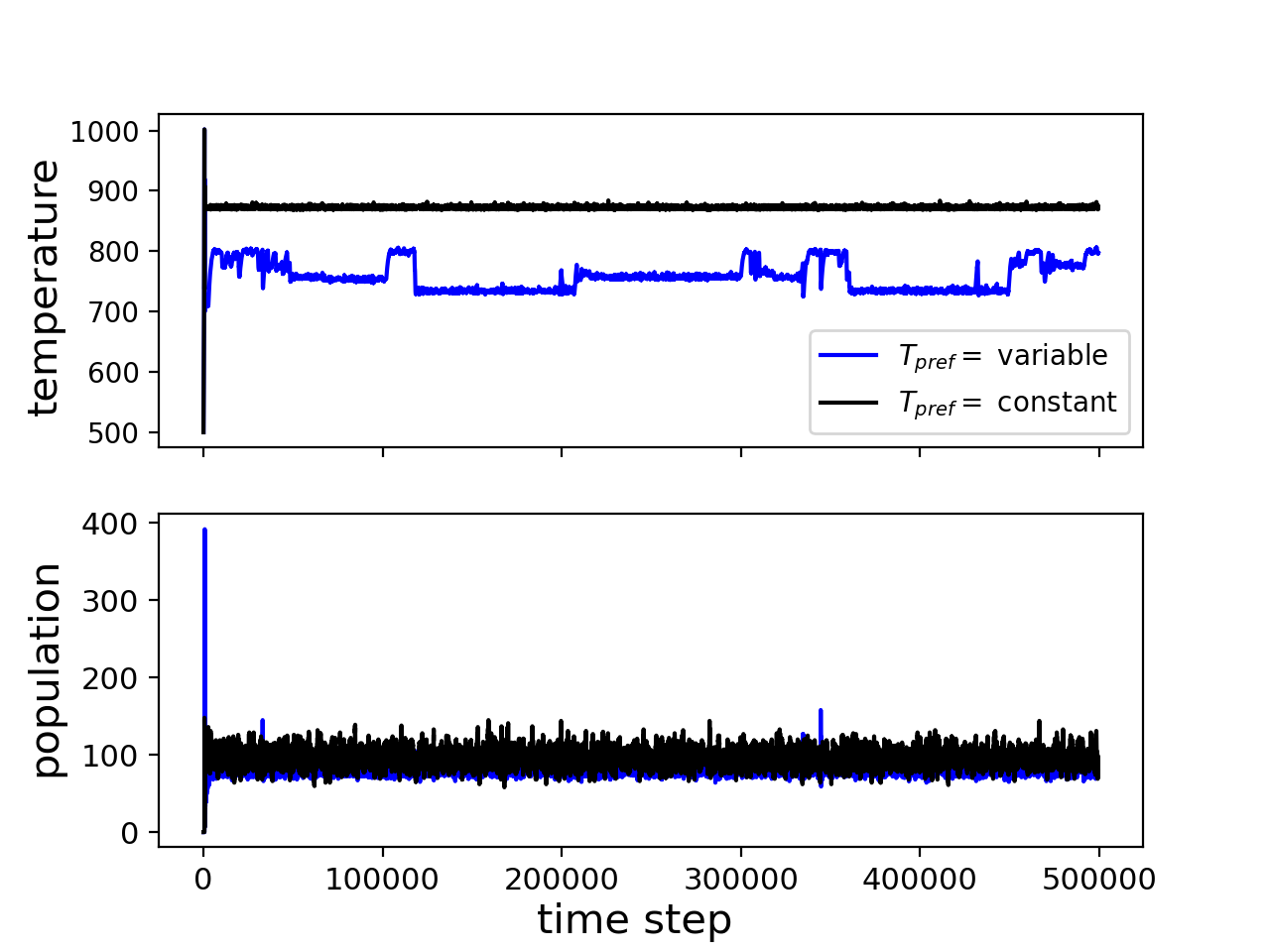}
  \caption{Long term behaviour Bottleneck}
  \label{fig:Bottleneck_example_b}
\end{subfigure}
\centering

\begin{subfigure}{.49\textwidth}
  \centering
  \includegraphics[scale=0.49]{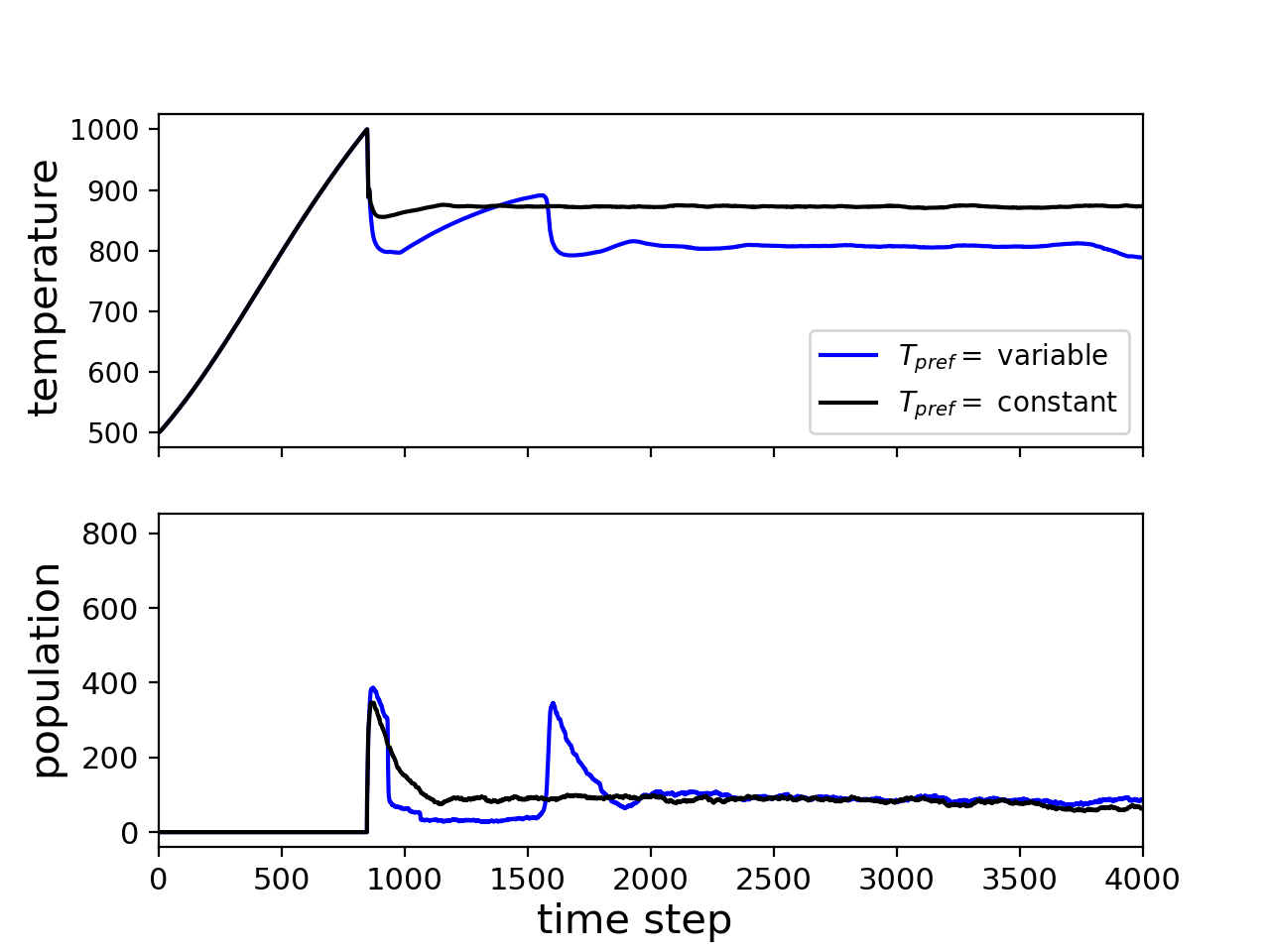}
  \caption{Behaviour soon after seeding Critical}
  \label{fig:Critical_example_a}
\end{subfigure}%
\begin{subfigure}{.49\textwidth}
  \centering
  \includegraphics[scale=0.49]{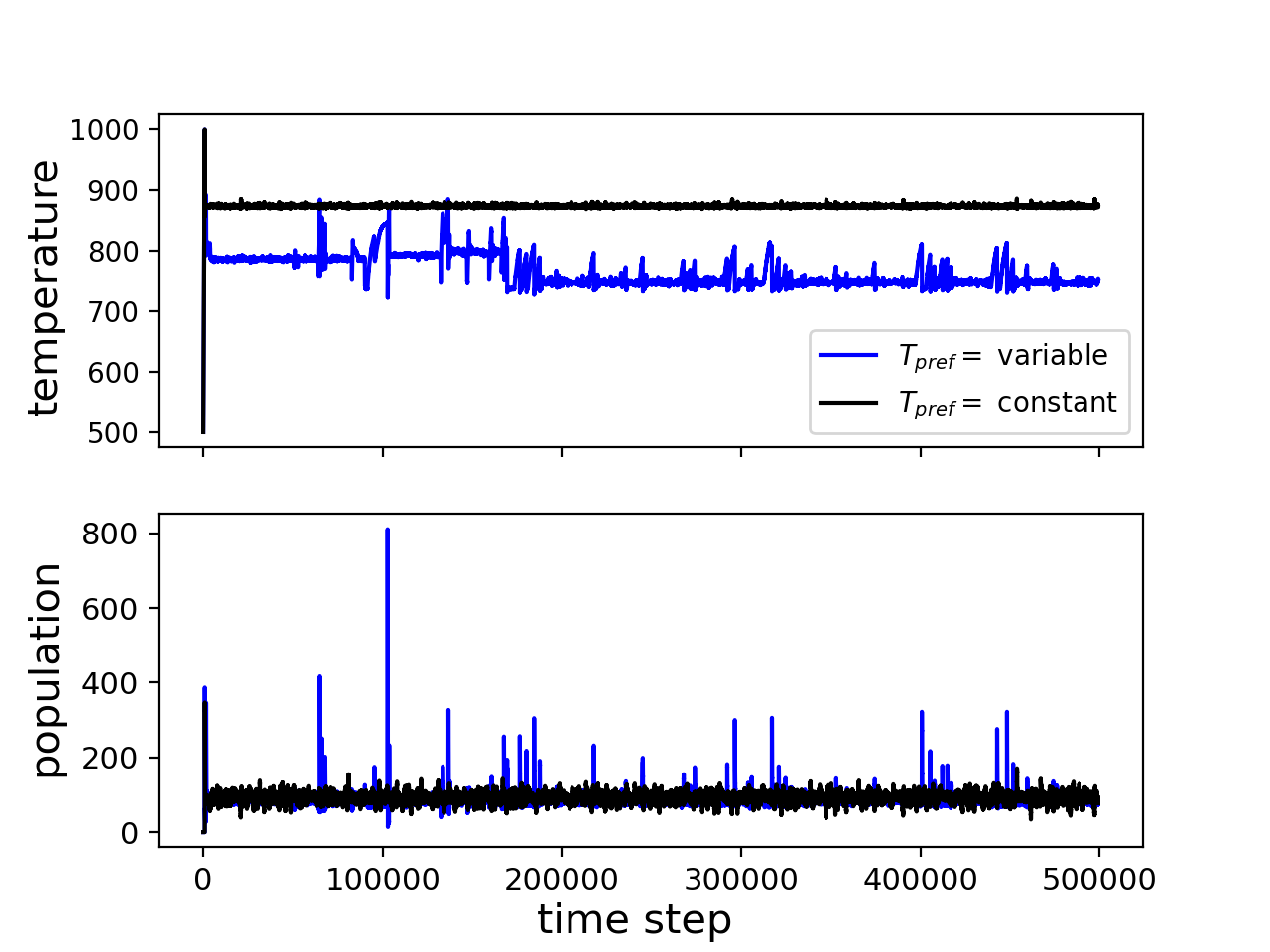}
  \caption{Long term behaviour Critical}
  \label{fig:Critical_example_b}
\end{subfigure}
\centering
\caption{Individual experiments for an Abiding, a Bottleneck and a Critical planet.}
\label{fig:Bottleneck_Critical_example}
\end{figure*}


Here we will present some individual experiments where we compare results from either keeping $T_{pref}$ constant for all microbe species and for allowing $T_{pref}$ to vary. In each case we choose to seed a single microbe species when $T = 1000$. When $T_{pref}$ is constant, all microbe species have peak growth rates when $T = T_{pref} = 1000$. For varying $T_{pref}$ we still seed when $T = 1000$ and dictate that the seeding species $T_{pref} = 1000$. Therefore the two experiments will start identically, and start to differ as mutation takes place. 

We chose examples from each classification of planet - as defined when $T_{pref}$ is constant in Section \ref{Section:hab_regulation_original}. We find that the survival rate of planets is impacted by allowing $T_{pref}$ to vary and thus planets that are classified as e.g. Abiding when $T_{pref}$ is fixed for all species, may suffer worse long term habitability prospects when $T_{pref}$ is allowed to vary. This is explored further in Section \ref{Section:survival_prospects}.

Figures \ref{fig:Abiding_example_a} and \ref{fig:Abiding_example_b} show the temperature and total population for two experiments for the same planetary set-up, one where $T_{pref}$ is constant for all species (shown in black) and one where $T_{pref}$ varies between species (shown in blue). This example planet is classified as `Abiding' when $T_{pref}$ is fixed, meaning that in all experiments, life survives until the end of the experiment.  We see where $T_{pref}$ is constant, the typical ExoGaia model behaviour where microbes `catch' a window of habitability, as they are seeded when $T = T_{pref}$ and after an initial population spike, the population stabilises, and the temperature is regulated at a near constant value where $T < T_{pref}$. When $T_{pref}$ varies, we find that the temperature drops to below that of the fixed $T_{pref}$ experiment, and evolves over time, experiencing sudden changes. The total population of the microbe community when $T_{pref}$ is variable is largely the same as when $T_{pref}$ is fixed, however where the temperature of the planet abruptly changes, the total population can spike, as seen at $t \approx 400,000$ in Figure \ref{fig:Abiding_example_b}.

%

Figures \ref{fig:Bottleneck_example_a} and \ref{fig:Bottleneck_example_b} show experiments for a Bottleneck planet (as classified for constant $T_{pref}$), and Figures \ref{fig:Critical_example_a} and \ref{fig:Critical_example_b} show these experiments on a Critical planet. Again for each each case we find that the temperature is regulated at a lower value when $T_{pref}$ varies between microbe species, and we see that the trajectory of the temperature and microbe population when $T_{pref}$ varies is noisier for the Critical planet (Figures \ref{fig:Critical_example_a} and \ref{fig:Critical_example_b}) . We see in Figure \ref{fig:Critical_example_b} that periods significant and rapid temperature fluctuations are associated with significant and rapid increases in the total microbe population.

\begin{figure*}
\centering
\begin{subfigure}{.49\textwidth}
  \centering
  \includegraphics[scale=0.49]{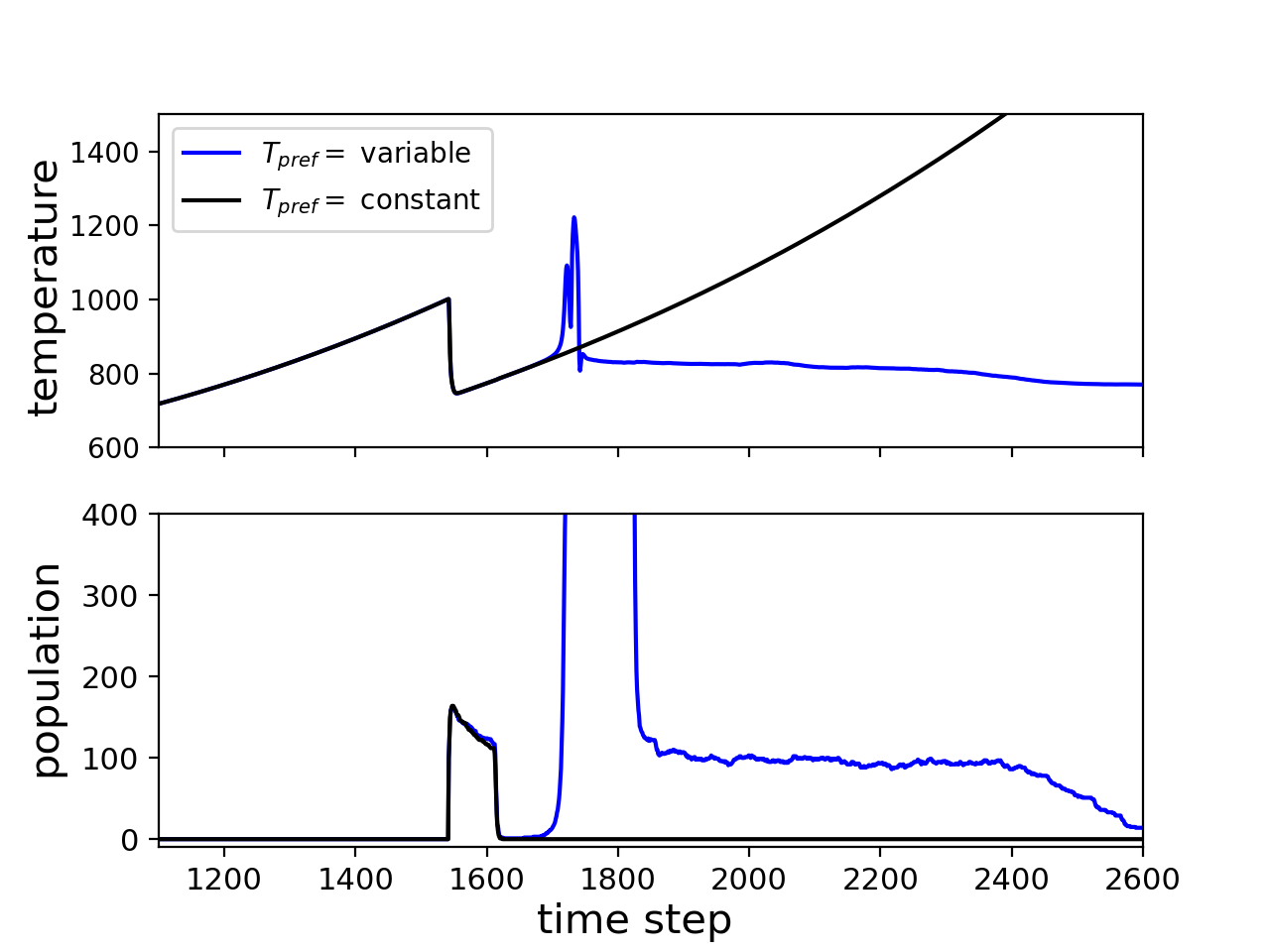}
  \caption{Behaviour soon after seeding Doomed planet where $T_{abiotic} >> T_{pref}$.}
  \label{fig:Doomed_example_a}
\end{subfigure}%
\begin{subfigure}{.49\textwidth}
  \centering
  \includegraphics[scale=0.49]{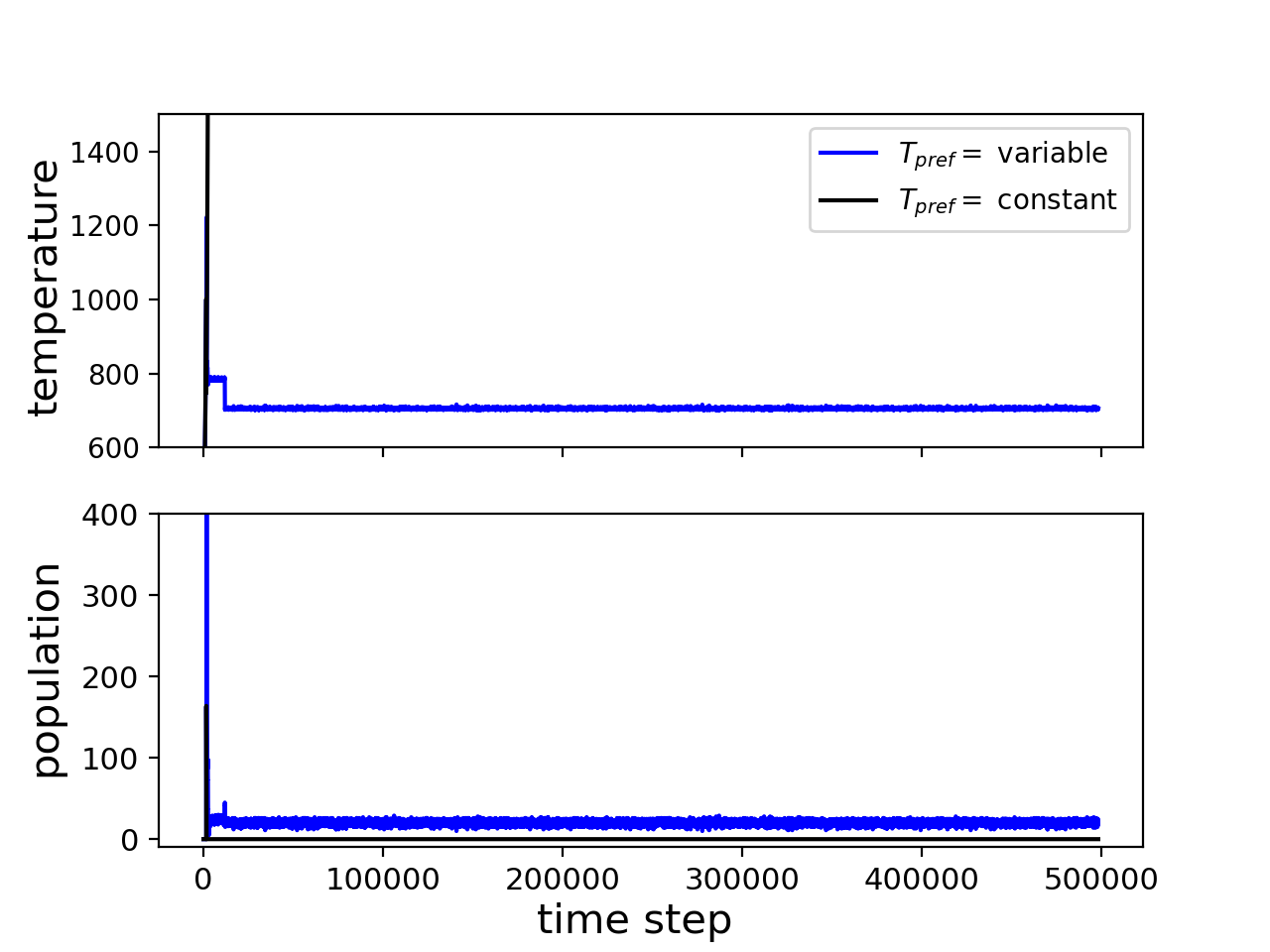}
  \caption{Long term behaviour Doomed planet where $T_{abiotic} >> T_{pref}$.}
  \label{fig:Doomed_example_b}
\end{subfigure}
\centering

\begin{subfigure}{.49\textwidth}
  \centering
  \includegraphics[scale=0.49]{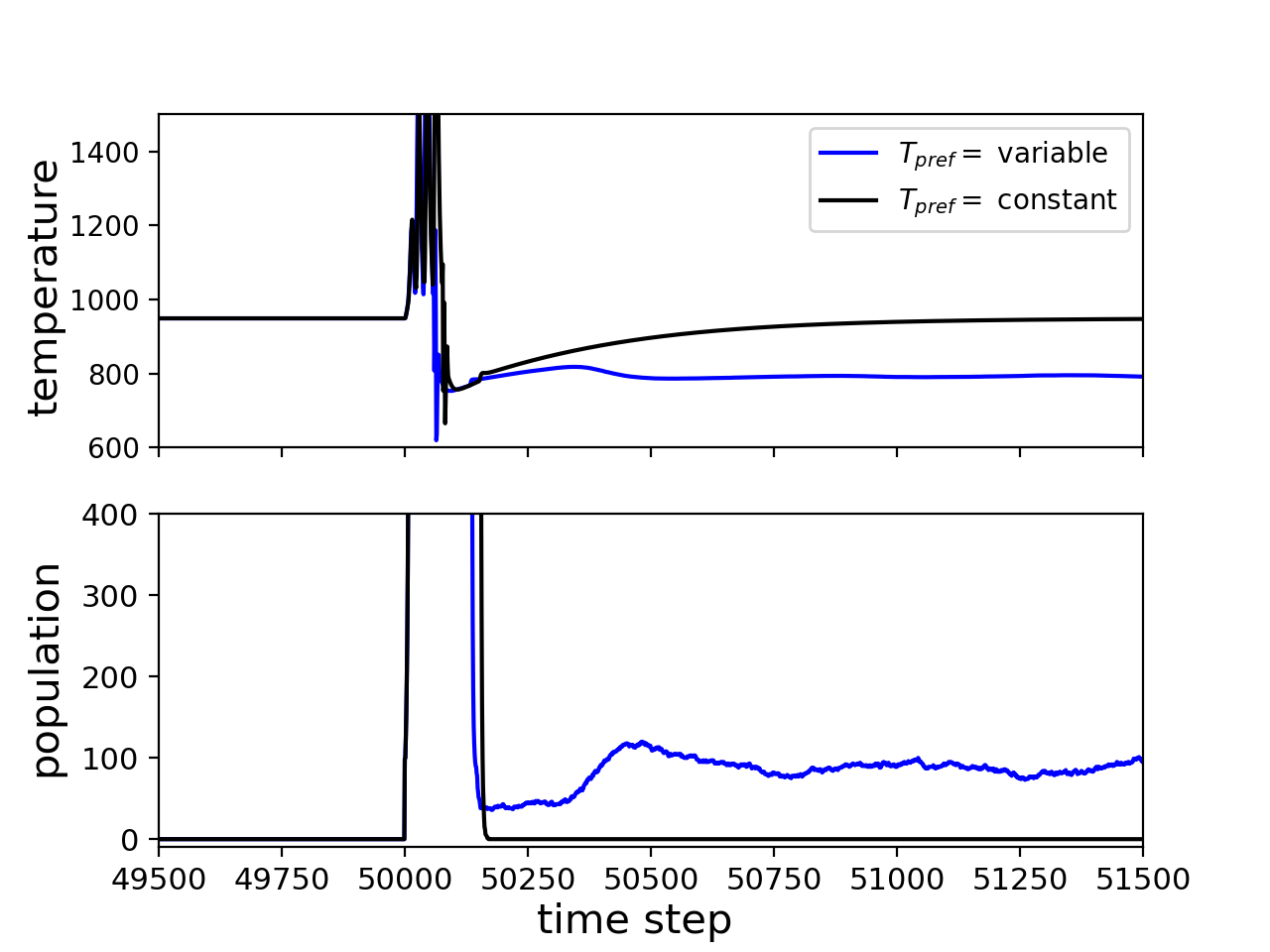}
  \caption{Behaviour soon after seeding Doomed planet where $T_{abiotic} \approx T_{pref}$}
  \label{fig:Doomed_example_c}
\end{subfigure}%
\begin{subfigure}{.49\textwidth}
  \centering
  \includegraphics[scale=0.49]{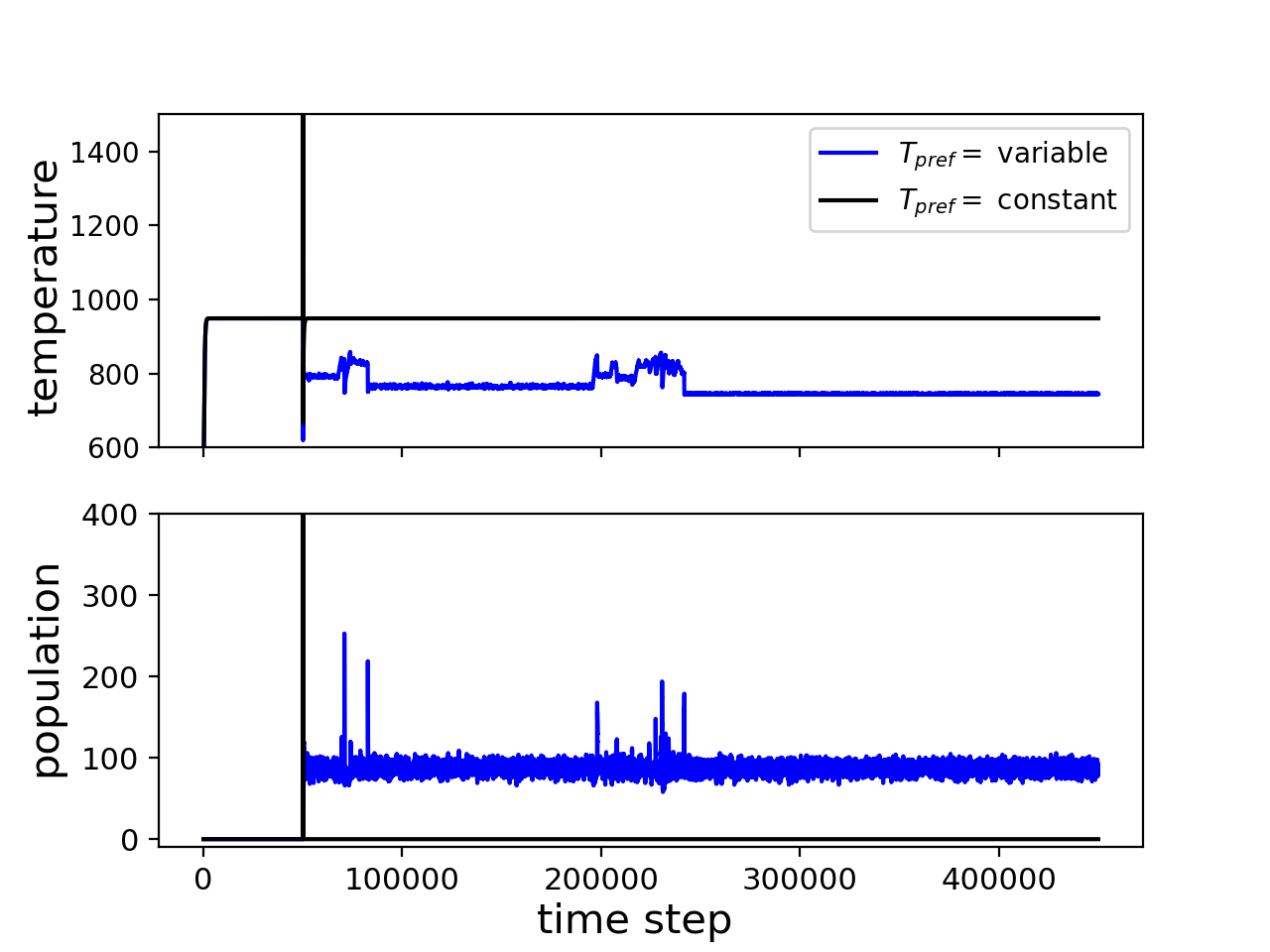}
  \caption{Long term behaviour Doomed planet where $T_{abiotic} \approx T_{pref}$}
  \label{fig:Doomed_example_d}
\end{subfigure}
\centering
\caption{Individual experiments for two different Doomed planets.}
\label{fig:Doomed_example}
\end{figure*}

For experiments where $T_{pref}$ is fixed for all microbe species a scenario can arise where the planet appears habitable - the temperature either passes through $T_{pref}$, or stabilises close to this value, yet life in every experiment is incapable of establishing itself on the planet. Where the abiotic temperature of a Doomed planet is such that $T_{abiotic} >> T_{pref}$, when life emerges on a planet, it cools the planet significantly and hinders its own growth causing a population drop. The abiotic background chemistry acts too slowly to improve conditions before all life goes extinct. Therefore the initial perturbation caused by life leads to its own extinction and establishment on the planet is not possible. For doomed planets where the abiotic temperature is such that $T_{abiotic} \approx T_{pref}$, microbial activity pushes the temperature outside the tolerable range as it changes the atmospheric composition and life is incapable of both consuming a sufficient amount of food to maintain a stable population, and keeping the temperature within habitable bounds. When $T_{pref}$ is allowed to vary the survival prospects of some of these Doomed planets changes. 

Figure \ref{fig:Doomed_example} shows an example from each type of Doomed planet, one where $T_{abiotic} >> T_{pref}$ and one where $T_{abiotic} \approx T_{pref}$. In each case allowing $T_{pref}$ to vary between species now allows the microbe community to establish temperature regulation successfully where communities with fixed $T_{pref}$ values failed. For the case where $T_{abiotic} >> T_{pref}$, shown in panels a) and b) in Figure \ref{fig:Doomed_example}, we see that the emergence of life on the planet, at around $T \approx 1550$, leads to a temperature drop to below that which the microbes with fixed $T_{pref}$ can survive, and although as the microbe population starts to drop the temperature begins to rise again, it does so slowly. In the fixed $T_{pref}$ experiment life therefore quickly goes extinct, however when $T_{pref}$ is varied for different species, microbes can emerge that can tolerate the new cooler conditions. This leads to a second population spike at roughly $t \approx 1800$ with corresponding fluctuations in the temperature. However both the population and temperature stabilise and habitability is now maintained for the full experiment. Doomed planets provide an interesting case where what might appear to be `ideal' abiotic conditions for life turn out to be less than ideal.

When $T_{abiotic} \approx T_{pref}$, we see in panels c) and d) in Figure \ref{fig:Doomed_example} that the initial introduction of life in both experiments causes a large population and corresponding rapid temperature increase. The temperature quickly drops, and when $T_{pref}$ is constant, the microbe community rapidly goes extinct. However when $T_{pref}$ varies, again microbes can adapt to the colder environment and establish and maintain habitable conditions until the end of the experiment. This example highlights that when conditions are `ideal' for life, as life interacts with and changes its environment it can degrade its own habitability. If however life is then able to adapt to the changing environment, this negative `anti-Gaian' (i.e. self-destructive) \citep{kirchner2002gaia} impact can be reduced. These examples demonstrate how the same feedbacks between life and its environment can be `Gaian' or `anti-Gaian' depending on the planetary context. The `Medea hypothesis' \citep{ward2009medea} has been proposed as a counter to the Gaia hypothesis suggesting that life is inherently self-destructive and draws on examples of mass extinctions in Earth history as evidence towards this. Doomed ExoGaia planets, where $T_{abiotic} \approx T_{ideal}$, might appear to be an example of this, however if life fails to become established on its planet, it is never fully integrated with its abiotic environment. For the majority of ExoGaia planets life must quickly integrate with its planet and establish regulatory feedback loops to prevent the planet reverting to an inhospitable state. This feature of the model is incompatible with a view that life is inherently self-destructive.

\subsection{Mean planetary temperature behaviour}
\label{section:MeanPlanetaryTempBehaviour}

The individual experiments presented in Section \ref{Section:individual_simulations} show examples where allowing $T_{pref}$ to vary between species leads to the biosphere cooling its planet to lower temperatures than for experiments where $T_{pref}$ is fixed. Here we will look at how the mean temperature of life-hosting ExoGaia planets is impacted by allowing $T_{pref}$ to vary between species.

\begin{figure}
\centering
\includegraphics[width=0.45\textwidth]{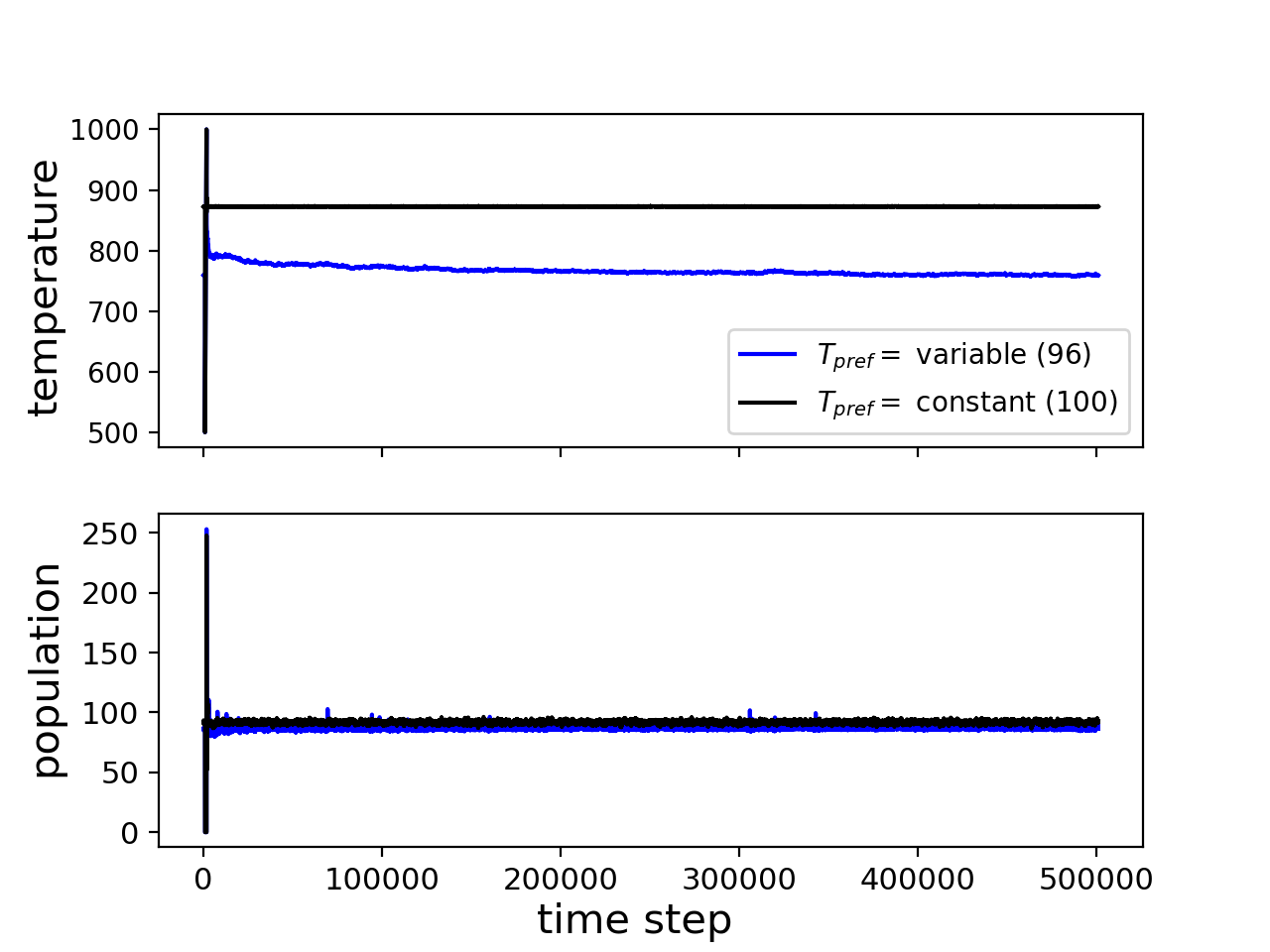}
\caption{The mean surface temperature, and mean microbe population for all surviving runs for a single ExoGaia planet setup. Black represents experiments where $T_{pref}$ is constant, and blue where $T_{pref}$ varies between microbe species. The numbers in brackets indicate the number of experiments averaged over in each case.}
\label{fig:geo2_average}
\end{figure}

Figure \ref{fig:geo2_average} shows the mean temperature and population over time for simulations where life survived until the end of the experiment (indicated by the number in brackets in the legend) on the same ExoGaia planet. 100 experiments were run for each scenario and when $T_{pref}$ was constant for all species, all 100 experiments had life surviving until the end. When $T_{pref}$ varied between species, 96 experiments out of 100 had life survive until the end of the experiment. Both cases - where $T_{pref}$ is constant, and where $T_{pref}$ varies between species - are shown in Figure \ref{fig:geo2_average}. We find that the mean temperature for planets hosting life where $T_{pref}$ varies is colder than for those hosting life where $T_{pref}$ is constant, and that the total populations of the microbial biospheres in each case are close in value.

Figure \ref{fig:geo2_average} shows that when $T_{pref}$ can vary, the mean behaviour is for the temperature of the ExoGaia planet to quickly cool to below the value it is regulated at when $T_{pref}$ is constant. While in an individual experiment, the temperature may at times increase (as seen in Figure \ref{fig:Bottleneck_example_b}) we find that the overall behaviour of the model is for the microbes to rapidly cool the planet to the lowest temperature a microbial biosphere can tolerate while successfully regulating their environment, and then stabilise at this value. 

Although there is nothing within our model specifically preventing microbes from regulating the temperature at temperatures above the original $T_{pref} = 1000$ we find that the mean behaviour is for the temperature to drop to below this value. This is due to the feedback loops between microbe metabolisms, and temperature. As described in Section \ref{Section:hab_regulation_original}, when $T_{pref}$ is fixed for all microbes, planetary regulation occurs only when $T < T_{pref}$. Where $T_{pref}$ differs for different microbe species, their growth rates will peak for different temperatures, some above the original $T_{pref} = 1000$, and some below. However the system tends to colder temperatures due to the same mechanism that lead to temperature regulation occurring only when $T < T_{pref}$ for the fixed $T_{pref}$ experiments. With differing $T_{pref}$ for different species, a `cascading' effect can take place pushing the temperature to lower and lower values. 

\begin{figure}
\centering
\includegraphics[width=0.45\textwidth]{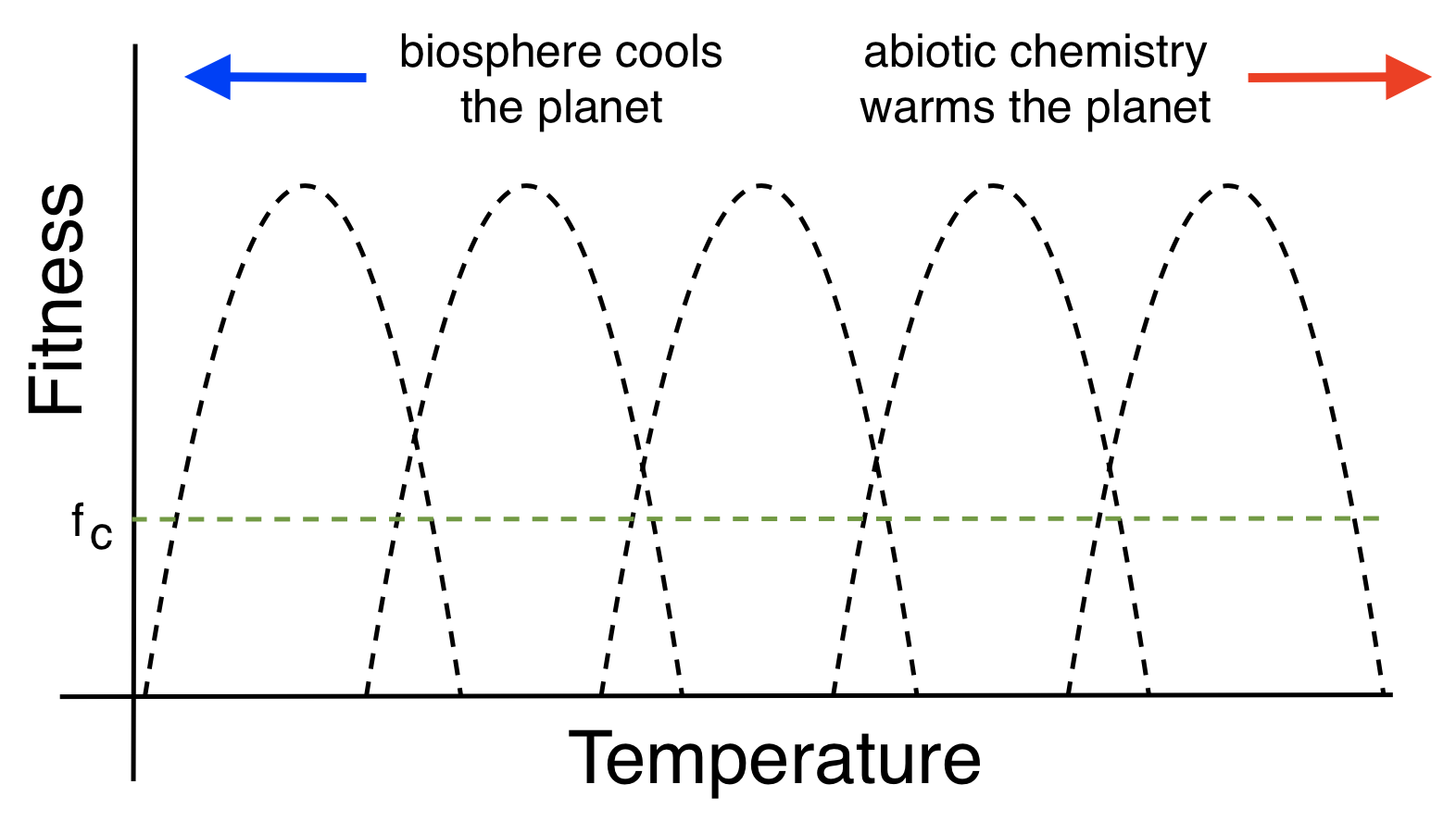}
\caption{Diagram showing varying temperature dependant fitness curved for different species when $T_{pref}$ varies. In each case, no matter the value of $T_{pref}$, the microbes act to cool the planet, whereas the background abiotic chemistry in the set ups explored act to warm the planet.}
\label{fig:different_T_pref}
\end{figure}

Figure \ref{fig:different_T_pref} shows a representation of multiple microbe species' fitness curves peaking for different temperatures. Each time a species evolves that has a slightly cooler $T_{pref}$ to the current microbe community, they push the temperature cooler due to the feedback between their growth rate and temperature as shown in Section \ref{Section:hab_regulation_original}. This pushes the climate to be too cold for the preexisting microbe community who then die out and are replaced by microbes that have a lower $T_{pref}$. Because the impact of a stable microbe biosphere on its environment always leads to global cooling due to the atmosphere being overall insulating, these same feedback loops do not occur when microbe species preferring higher temperatures mutate into existence. Therefore ExoGaia planets hosting microbial life with diverse temperature preferences will tend to quickly cool to the lowest temperature a microbe community can survive at and still be able to successfully regulate their environment.

The abiotic state for the majority of potentially habitable ExoGaia planets is that of an inhospitably hot planet, and so we can think of the biosphere as in effect preventing a runaway greenhouse on their planet. When allowing the microbial biosphere the ability to adapt to different planetary temperatures, we might expect that as the abiotic state of the planet is incredibly hot that the microbe community would adapt towards hotter climates. Instead the opposite behaviour emerges, where the microbial community instead further cool their planet due to a `cascade' of overlapping feedback loops between planetary temperature and the growth rate of different species.

\subsection{Survival prospects}
\label{Section:survival_prospects}

Here we investigate how allowing $T_{pref}$ to vary between species impacts the success rate for establishment of life and maintenance of long term habitability on ExoGaia planets. 

As described in Section \ref{Section:hab_regulation_original} the planet classifications used to describe ExoGaia planets are based on two time scales - establishment and long term habitability. Introducing varying $T_{pref}$ introduces a new source of internal perturbations to the system. However with varying $T_{pref}$ the biosphere is able to adapt to fluctuations in the planetary temperature, something not possible when $T_{pref}$ is fixed. We find that whether the habitability prospects of a planet are improved or degraded by allowing varying $T_{pref}$ depends on the planetary context and at what time in the experiment we are observing the planet.

\begin{figure}
\centering
\includegraphics[width=0.45\textwidth]{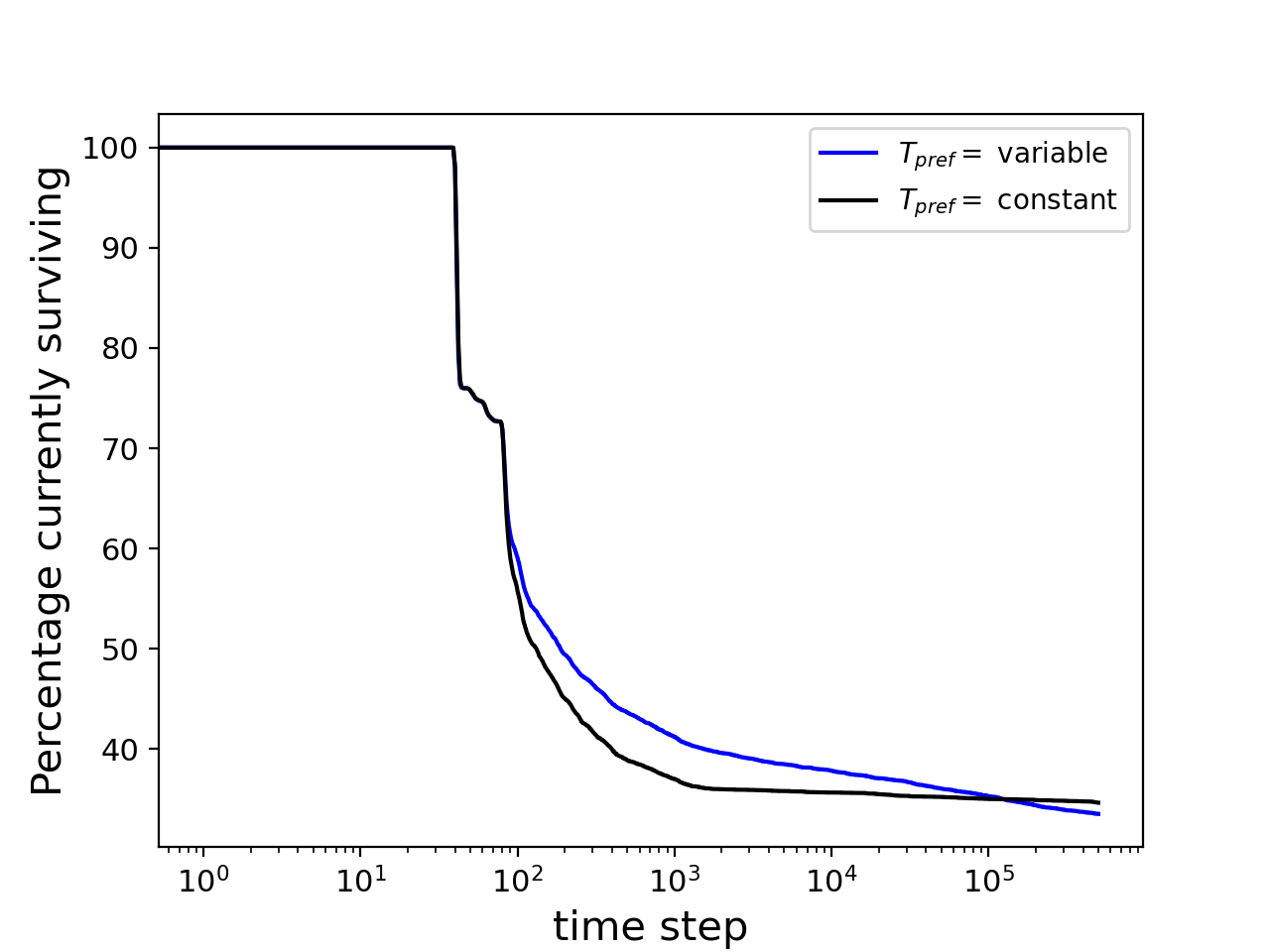}
\caption{Figure showing the percentage of planets surviving at every time step. Data includes all 10,000 simulations (100 experiments performed for each 100 planets) explored in this work.}
\label{fig:percent_survivng}
\end{figure}

Figure \ref{fig:percent_survivng} shows the percentage of experiments (out of all 10,000 experiments - 100 experiments on 100 planets) that have life present for each time step. Data is shown for both sets of experiments - where $T_{pref}$ is constant for all species, and where $T_{pref}$ varies between species. Note the logged x axis. Figure \ref{fig:percent_survivng} shows a rapid drop off early on as life that emerges on a planet incompatible with its survival rapidly goes extinct, and is followed by a slower decline at later times. We see that for early time, experiments allowing $T_{pref}$ have a higher survival rate than for experiments where $T_{pref}$ is fixed. Over time however, this improvement in survival rate for varying $T_{pref}$ reduces, until towards the end of the experiment, allowing $T_{pref}$ to vary results in a lower survival rate.

From Figure \ref{fig:percent_survivng} we see that overall, it is slightly beneficial for the microbial biosphere to be able to adapt to temperature changes early on when they are becoming established on a planet, but as time goes on this feature hurts their long term survival rate. When life first emerges on a planet it often causes a large perturbation as the microbes first begin to interact with their environment (see the figures for individual experiments in Section \ref{Section:individual_simulations}), therefore at early times, soon after emerging, the ability to adapt to large temperature fluctuations (e.g. the ability for microbes to emerge that can tolerate the new temperature) can help the biosphere survive such perturbations and avoid extinction. Introducing varying $T_{pref}$ however also introduces an additional source of perturbations, as the microbe community changes the value at which the planetary temperature is regulated at can change, and this additional source of perturbation can, over time, hinder long term survival prospects for a biosphere.

The original classification of the planet, e.g. what classification the planet would have for a biosphere where $T_{pref}$ is constant, plays a role in determining how beneficial or not introducing varying $T_{pref}$ is to the establishment and the long term habitability prospects of life on that planet. Table \ref{tab:planet_abundances} shows the abundance of each planetary classification under scenarios where $T_{pref}$ is constant for the 100 planetary set ups explored in this work. We can separate the data for both sets of experiments using the planet classifications in Table \ref{tab:planet_abundances} to determine the differences in behaviour between different planet classifications when introducing varying $T_{pref}$. 

\begin{table}
    \centering
    \begin{tabular}{c|c} \hline
    Planet classification & Frequency \\ \hline \hline
       Abiding  & 12 \\ \hline
       Bottleneck  & 27  \\ \hline
       Critical & 21 \\ \hline
       Doomed & 10 \\ \hline
       Extreme & 30 \\ \hline
    \end{tabular}
    \caption{Table showing the frequency of each planet classification for experiments where $T_{pref}$ is constant for all microbial species. Classifications for a planet are determined by analysing the results of 100 individual experiments with the same abiotic planetary configuration.}
    \label{tab:planet_abundances}
\end{table}

\begin{figure}
\centering
\includegraphics[width=0.45\textwidth]{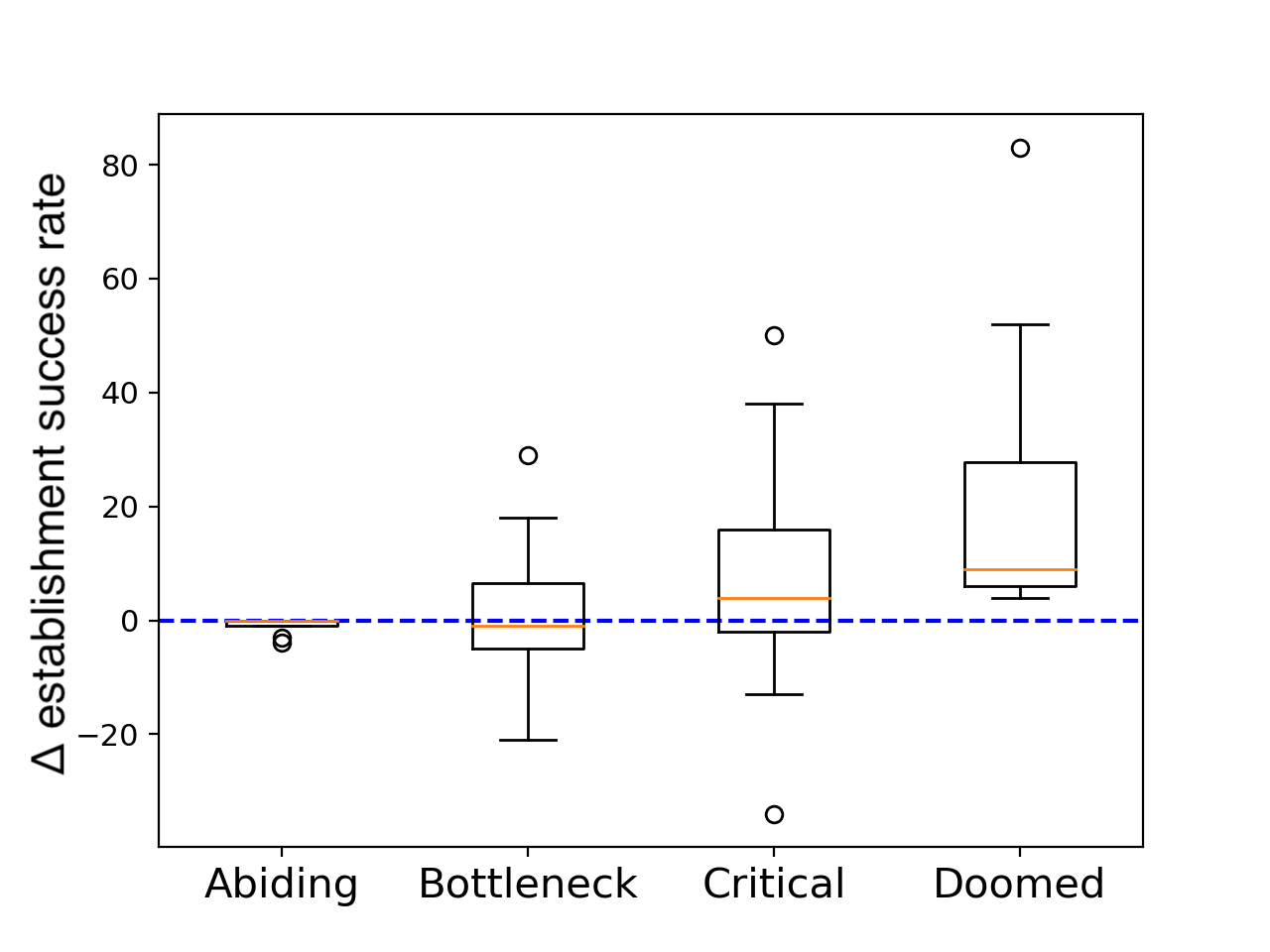}
\caption{Change in establishment success rate when allowing $T_{pref}$ to vary between species.}
\label{fig:establishment_change}
\end{figure}

In Figure \ref{fig:establishment_change} shows whisker and box plots \footnote{Created using the matplotlib.boxplot function in Python.} for the change in establishment rate (out of 100 experiments for each planet) when we introduce varying $T_{pref}$ for each planetary class (as defined when $T_{pref}$ is constant). The orange line in each box plot indicates the median change in establishment rate. The edges of the box extend from the first quartile to the third quartile, and the whiskers of the box extend to 1.5 times the interquartile range. Outliers are indicated by individual points. If a planet initially had a success rate of $50 / 100$ and this changes to $20 / 100$ when introducing varying $T_{pref}$ this would be recorded as a change of $-30$. These changes in establishment success rates are calculated for each planet and grouped via planet classification as defined when $T_{pref}$ is fixed. Extreme planets are not included in Figure \ref{fig:establishment_change} as they have conditions incompatible with life establishment under both experimental scenarios. 

We find that the establishment rate of Abiding planets is minimally impacted. By definition, when $T_{pref}$ is fixed, Abiding planets have a establishment rate of $100\%$ and so this rate cannot be improved with introducing varying $T_{pref}$ but could be negatively impacted. We find however that the establishment success for Abiding planets is only very minimally impacted by introducing varying $T_{pref}$.

For Bottleneck planets, Figure \ref{fig:establishment_change} shows a median value of the change in establishment success that is very close to zero, with the box edges and whiskers positioned roughly symmetrically about zero. This indicates that the overall establishment success is unchanged for Bottleneck planets, but that individual experiments on Bottleneck planets can be significantly impacted, both positively and negatively, by allowing $T_{pref}$ to vary. On a Bottleneck planet, when $T_{pref}$ is constant, the emergence of life causes a large perturbation on the planet that life may or may not survive. Allowing $T_{pref}$ to vary can allow for life to adapt to this changing temperature and so avoid extinction, but can also introduce an additional source of perturbation into the system which could drive the biosphere to extinction.

For Critical planets Figure \ref{fig:establishment_change} shows a slight improvement on the overall establishment success rate of planets, but again we see a large variance in the data this time more skewed towards positive values. Critical planets are prone to perturbations even when hosting microbes where $T_{pref}$ is fixed, and Figure \ref{fig:establishment_change} demonstrates that with varying $T_{pref}$ where microbes have a greater ability to adapt to climactic changes occurring on their planet the establishment success is slightly higher.

Figure \ref{fig:establishment_change} shows that Doomed planets experience clear improvement in establishment success when we allow $T_{pref}$ to vary. When $T_{pref}$ is constant, these planets have a zero success rate of establishment as the initial perturbation of life emerging on the planet always leads to total extinction as conditions recover too slowly, or not sufficiently enough to allow for life to survive (see Figure \ref{fig:Doomed_example}). On these planets, the ability of life to adapt to changing temperatures allows for cases where life can now survive the initial perturbation, and are able to interact with their planet sufficiently to survive and adapt to the resulting temperature. 

Overall Figure \ref{fig:establishment_change} demonstrates that not only is the establishment success for microbe communities where $T_{pref}$ can vary higher than that for where this is fixed, agreeing with Figure \ref{fig:percent_survivng} but that when we analyse each planet class, as defined when $T_{pref}$ is fixed, no planetary class experiences a significantly decreased rate of establishment success when $T_{pref}$ is allowed to vary between species.

\begin{figure}
\centering
\includegraphics[width=0.45\textwidth]{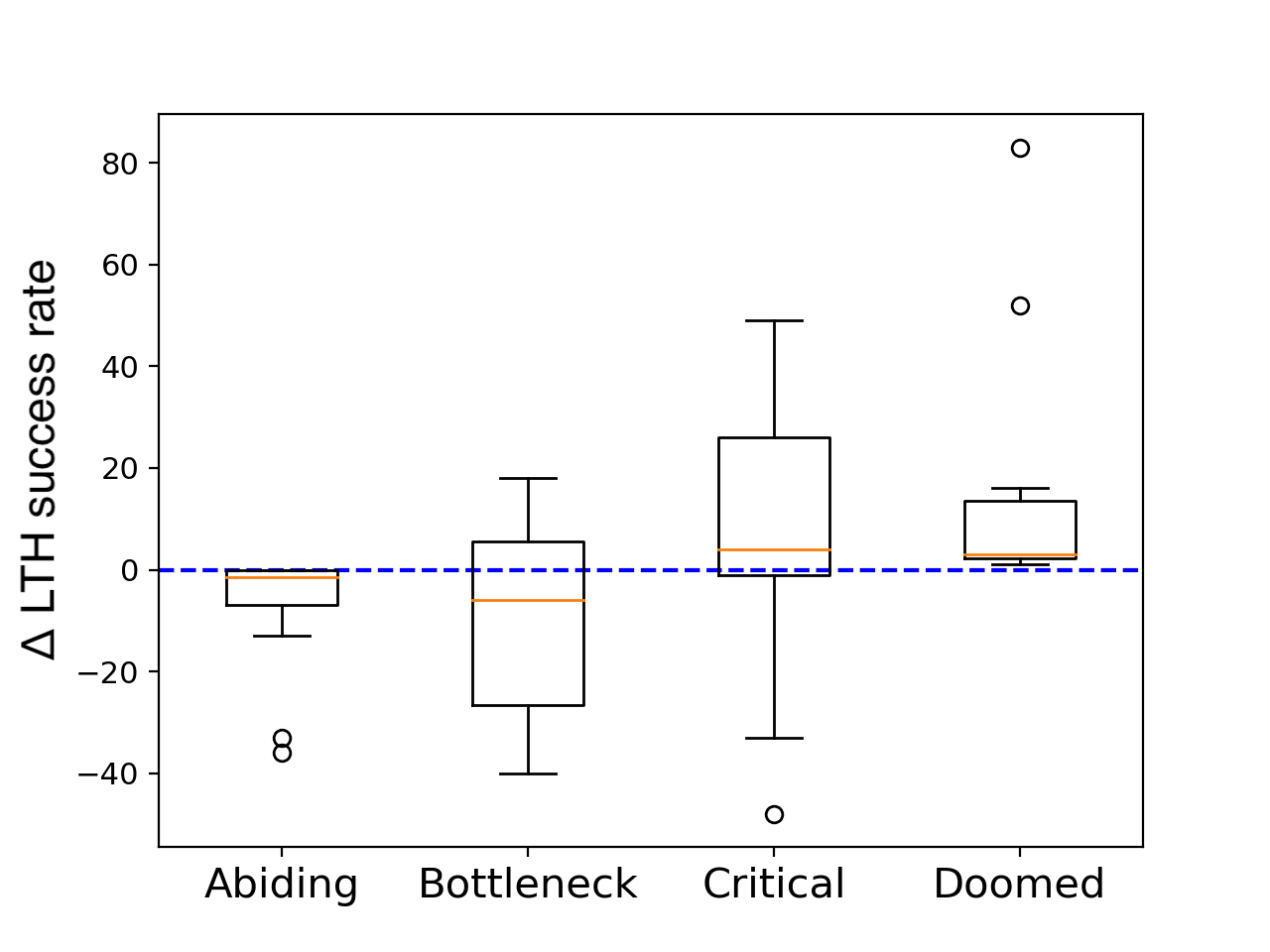}
\caption{Change in long term habitability success rate when allowing $T_{pref}$ to vary between species.}
\label{fig:LTH}
\end{figure}

Figure \ref{fig:LTH} shows whisker and box plots for the change in long term habitability success rate (out of 100 experiments for each planet). Note that the same format for these plots applies as in Figure \ref{fig:establishment_change}. These rates are calculated as for the establishment success rates, but instead for the number of planets that successfully host life until the end of the experiment. We find that the planet classification impacts whether introducing varying $T_{pref}$ improves or hinders the overall long term survival rate for life on a planet.

Figure \ref{fig:LTH} shows that the long term habitability (LTH) success rates of Abiding planets is only slightly negatively impacted by introducing varying $T_{pref}$. The median value for the change in success rate (indicated by the orange line) is close to zero, although some outliers show a significant decrease in success rate. Overall introducing varying $T_{pref}$ does not significantly impact the establishment success rate, and only minimally negatively impacts the overall long term habitability of Abiding planets. Abiding planets have background abotic chemistry that form recycling feedback loops between atmospheric chemicals even in the absence of life (which adds additional feedback loops via its metabolism). This makes them less prone to perturbations and overall provide a more stable environment for life to interact with and regulate, and aids in the rapid recovery of habitable conditions after perturbations. Therefore the introduction of additional perturbations by letting $T_{pref}$ vary overall does not greatly impact this stability. While some individual Abiding planets can have their LTH prospects significantly reduced by allowing $T_{pref}$ to vary, for the majority this change in the biosphere has little impact on establishment and LTH success rates. 

Bottleneck planets in Figure \ref{fig:LTH} show an overall decrease in their long term habitability success rate. The median change in LTH success rate is negative and the data skews towards negative values. On a Bottleneck planet when $T_{pref}$ is fixed, introducing life to the planet results in a large perturbation. If life survives this initial period and successfully establishes the planet then it goes on to survive the full experiment. Therefore this initial perturbation is the largest `challenge' for life to overcome on Bottleneck planets. When allowing $T_{pref}$ to vary, an additional source of perturbations is introduced which can present additional challenges for life to overcome. In the case of a Bottleneck planet where after successful establishment, if $T_{pref}$ is fixed, LTH success is guaranteed, this additional source of perturbations overall negatively impacts the LTH changes for the planet. Bottleneck planets that experience a higher LTH success rate with varying $T_{pref}$ are those that had a higher establishment success rate with varying $T_{pref}$ enough to offset the additional source of perturbation throughout the experiment. 

Figure \ref{fig:LTH} shows an overall increase in the LTH success rate for Critical planets. The median data point is slightly positive, and the data skews towards positive values. Critical planets are prone to extinction causing perturbations at any point of their history as the background abiotic chemistry contains fewer preexisting recycling loops. The microbe community must in effect `fill in' any missing links between atmospheric chemicals to form recycling loops, but as microbe metabolisms are temperature sensitive these links are prone to fluctuations that can cause the recycling of atmospheric chemicals to break down, leading to the uncontrolled build up of atmospheric chemicals which can ultimately lead to extinction. Allowing $T_{pref}$ to vary introduces additional perturbations to the system, but also allows the microbe community to adapt to changing conditions and ultimately Figure \ref{fig:LTH} demonstrates that this leads to an increase in the overall LTH success rate for Critical planets. We see again that there is a large variation in the change in LTH success rate for Critical planets, some are strongly negatively impacted, and others are strongly positively impacted. Therefore while overall LTH success rates are improved, an individual experiment for a Critical planet might suffer an extinction causing perturbation by allowing $T_{pref}$ to vary.

Doomed planets would have a establishment success rate of zero with fixed $T_{pref}$ and therefore a LTH success rate of zero. Figure \ref{fig:LTH} shows that allowing $T_{pref}$ to vary leads to an increase in the overall LTH success rate of Doomed planets, and that for some individual Doomed planets their LTH success rates are significantly improved. Therefore for a Doomed planet, it is beneficial for the microbe community to have differing $T_{pref}$ values for different species. 

Overall, the largest decline in long term habitability prospects due to varying $T_{pref}$ between species is seen for Bottleneck planets, and the largest improvement is seen for Critical planets. The results shown in Figure \ref{fig:establishment_change} and \ref{fig:LTH} further demonstrate that the same actions by life interacting with its planet could be deemed either `anti-Gaian' or `Gaian' depending on the planetary context and the particular trajectory of any experiment. These results further the hypothesis that the long term habitability prospects for a planet is not a deterministic feature of planet, but a statistical phenomena that emerges from the interactions between life and the abiotic environment, and that individual planets could have many possible climatic trajectories that they follow \citep{lenardic2016solar, chopra2016case, lenardic2019toward}.

\section{Conclusions}
\label{discussion}

We have extended the ExoGaia model to investigate how allowing the `ideal' temperature ($T_{pref}$) for microbes to vary between species impacts the regulation of the surface temperature of ExoGaia planets, and how this change impacts both the establishment success, and the LTH success of planets. For the model setup explored the majority of potentially habitable ExoGaia planets have abiotic temperatures that are far too high for life to tolerate. However when a microbial biosphere is introduced to the planet, it is capable of `catching' a window of habitability and maintaining it for long time spans. By consuming parts of the overall insulating atmosphere the collective impact of the biosphere is to cool its host planet.

We find that, overall, when allowing temperature preferences to vary between species, the microbial community drive the climate to lower temperatures, towards the limit of what they can tolerate while still being able to successfully recycle atmospheric chemicals and maintain a constant population. This is due to the default setup of ExoGaia planets explored in this work, where the atmosphere is overall insulating, and so the impact of the biosphere on the atmosphere is to reduce its insulating effect, as it removes atmospheric chemicals. This leads to the microbe community regulating at lower temperatures via a `cascade' effect when $T_{pref}$ varies between species - the temperature is driven to lower and lower temperatures as microbes evolve that are better adapted to cooler climates (see Section \ref{section:MeanPlanetaryTempBehaviour} for the mechanism behind this behaviour). 
Therefore, to form biosignature predictions for a temperature limited biosphere these results would imply that the strength of the biosignature produced would be impacted by the point at which the biospheres' impact on the climate becomes self limiting to its growth.

These results predict that inhabited planets with insulating atmospheres might have significantly colder climates than abiotic models would predict, as is the case on Earth \citep{schwartzman1989biotic}. Allowing temperature adaptation of the microbe community only accentuates the difference between the biotic and abiotic state of an ExoGaia planet. The overall behaviour of the model is not for the biosphere to adapt towards the abiotic state of a warm climate, but for the planetary temperature to be kept cool via the biosphere's regulation of the atmosphere. This feedback mechanism, where a temperature limited biosphere regulates the climate, is proposed to be a method for stabilising the climate of potentially habitable $H_{2}$ dominated greenhouse planets \citep{abbot2015proposal} thus maintaining habitability on these planets for much longer time spans. Therefore understanding the behaviour of a temperature limited biosphere might be important for understanding the habitable zone, and long term habitability prospects, for such planets.

The behaviour of the ExoGaia model predicts that observational evidence of two planets at similar radii around similar stars, but with significantly different climates, might indicate the presence of a biosphere on one planet but not the other. It is speculated that in our own solar system the dramatically different climates of Venus and Earth, despite similar bulk properties, might not solely be the results of their differing levels of incoming radiation from our sun, but instead might instead represent two alternative states of the same planetary system \citep{lenardic2016solar}. If this is the case, and if life can play a role in determining the end climactic state of its planet, then it could be possible that an inhabited Venus would be significantly cooler than Venus' current climate. While we cannot `rewind' time to the beginning of our solar system to investigate possible alternative states for Earth and Venus, using observational data from other solar systems similar to our own gives us the opportunity to test whether Venus and Earth's present day climates are more, or less, deterministic. Finding a solar system similar to our own, but with an Earth-like planet closer in towards the star and a Venus-like planet further out could indicate that the climatic states of these planets are not solely due to positioning in the solar system but to additional factors, including possibly the presence, or lack thereof, of a biosphere. 

While ExoGaia is an abstract model and so cannot be used for forming quantitative predictions for specific planets, the results explored in this work mirror some of the behaviour regarding how life has impacted the Earth over time. It is predicted that the Earth would be significantly warmer in the absence of life \citep{schwartzman1989biotic} and it is thought that during Earth's history life has triggered significant periods of glaciation \citep{kopp2005paleoproterozoic}. That the ExoGaia model reproduces some of the climactic behaviour of Earth's history, albeit in a highly idealised and abstract way, increases our confidence in using the model to form generalised predictions for the search for life beyond our own planet.

These results demonstrate that the same impacts from life on its planet could be beneficial or harmful to its habitability prospects, and that this depends on the planetary context. Therefore the presence of `anti-Gaian' impacts of life on its environment is not incompatible with Gaia theory but further demonstrates that the whole system - both the abiotic planetary components and the biosphere must be considered together to form a complete understanding of the planet. Doomed planets in the ExoGaia model demonstrate that what could appear to be a hospitable planet under abiotic conditions, does not necessarily translate to a high rate of establishment success by life, or long term habitability. As life interacts with its environment it will change it, and if conditions were already `ideal' this impact by life will degrade the environment and could push conditions to beyond habitable bounds. If life is able to adapt to changing conditions however the risk of extinction is reduced. 

Our results show that overall the establishment success of ExoGaia planets is improved by allowing the `ideal' temperature preferences to vary between species, however the long term habitability (LTH) success rates are slightly decreased, most significantly decreased for Bottleneck planets. However the life in ExoGaia is incredibly simple, and there is no scope within the model for increasing ecosystem complexity (e.g. the evolution of decomposers or secondary consumers). Multiple models demonstrate that increasing complexity of ecosystems leads to a higher level of stability \citep{harding1999food, christensen2002tangled, becker2014evolution, gaianhabzone}, meaning that over time as a biosphere becomes more complex it would become more stable and less susceptible to perturbations. Additionally microbes can become more resilient to perturbations if they have previously experienced similar stressors in their history \citep{lambert2014memory}. These features of more complex ecosystems could counteract the slight decrease in LTH success seen in the ExoGaia model when $T_{pref}$ varies between microbe species. The ExoGaia model is most representative of simple life emerging on a young planet before enough time has sufficiently elapsed for the biosphere to become more complex, and we find that the ability of microbes to adapt to changing environmental conditions overall improves their changes of success for successfully becoming established on their planet.

These results add to hypotheses that the `habitable zone' around a star might not be a deterministic property of a given stellar system with sharp boundaries, but rather a statistical probability of finding a life bearing planet \citep{gaianhabzone}.
This work takes steps towards beginning to determine a probabilistic biotic habitable zone.

\section{Future work}
\label{Section:further_work}

As visiting a distant planet to directly observe life is not possible, detailed abiotic and biotic models of planetary climates are crucial for verifying potential future biosignature detections \citep{catlingdavid2018exoplanet, walker2018exoplanet}. Determining whether a potential biosignature is indicitive of life will strongly depend on the context of the planet being observed \citep{Seager:2013a, Claudi:2017, Kiang:2018, Schwieterman:2018, Krissansen-Totton:2022}.

Forming predictions for biosignature observations requires detailed realistic models of life-climate interaction for a wide array of planetary conditions \citep{krissansen2022understanding}. Detailed abiotic exoplanet models are being developed for a wide range of detected planets \citep{amundsen2016uk, Boutle:2017, collins2021modeling, fauchez2021trappist}, and models of biogeochemistry have been developed for various points in Earth's history \citep{Kharecha:2005, Daines:2017, Lenton:2018, zakem:2020}. Combining these modelling efforts will be crucial for forming biosignature predictions to compare to real data from distant planets.

\cite{nicholson2022predicting} takes a step towards developing a framework for predicting biosignatures for life limited by nutrient availability, and a biomass based model has been developed by \cite{seager2013biomass} to estimate the plausibility of biosignature gases. Similar work is required to extend this work into developing realistic models of potential biosignatures for life limited by planetary temperature. With these different models for assessing the potential biospheres for life under different limiting regimes, we can develop a suite of biosignature predictions to compare with observational data from exoplanets.

\section*{acknowledgements}
This work was partly funded by the Leverhulme Trust through a research project grant [RPG-2020-82] and a UKRI Future Leaders Fellowship [MR/T040866/1].

We thank Aditya Chopra for his valuable feedback on this paper.

\section*{Data Availability}

The code is available on request from the authors.

\bibliographystyle{mnras}
\bibliography{references}

\begin{thebibliography}{}
\makeatletter
\relax
\def\mn@urlcharsother{\let\do\@makeother \do\$\do\&\do\#\do\^\do\_\do\%\do\~}
\def\mn@doi{\begingroup\mn@urlcharsother \@ifnextchar [ {\mn@doi@}
  {\mn@doi@[]}}
\def\mn@doi@[#1]#2{\def\@tempa{#1}\ifx\@tempa\@empty \href
  {http://dx.doi.org/#2} {doi:#2}\else \href {http://dx.doi.org/#2} {#1}\fi
  \endgroup}
\def\mn@eprint#1#2{\mn@eprint@#1:#2::\@nil}
\def\mn@eprint@arXiv#1{\href {http://arxiv.org/abs/#1} {{\tt arXiv:#1}}}
\def\mn@eprint@dblp#1{\href {http://dblp.uni-trier.de/rec/bibtex/#1.xml}
  {dblp:#1}}
\def\mn@eprint@#1:#2:#3:#4\@nil{\def\@tempa {#1}\def\@tempb {#2}\def\@tempc
  {#3}\ifx \@tempc \@empty \let \@tempc \@tempb \let \@tempb \@tempa \fi \ifx
  \@tempb \@empty \def\@tempb {arXiv}\fi \@ifundefined
  {mn@eprint@\@tempb}{\@tempb:\@tempc}{\expandafter \expandafter \csname
  mn@eprint@\@tempb\endcsname \expandafter{\@tempc}}}

\bibitem[\protect\citeauthoryear{Abbot}{Abbot}{2015}]{abbot2015proposal}
Abbot D.~S.,  2015, The Astrophysical Journal Letters, 815, L3

\bibitem[\protect\citeauthoryear{Abe, Abe-Ouchi, Sleep  \& Zahnle}{Abe
  et~al.}{2011}]{abe2011habitable}
Abe Y.,  Abe-Ouchi A.,  Sleep N.~H.,   Zahnle K.~J.,  2011, Astrobiology, 11,
  443

\bibitem[\protect\citeauthoryear{{Ahrer} et~al.,}{{Ahrer}
  et~al.}{2022}]{ahrerERS2022}
{Ahrer} E.-M.,  et~al., 2022, arXiv e-prints, \href
  {https://ui.adsabs.harvard.edu/abs/2022arXiv221110489A} {p. arXiv:2211.10489}

\bibitem[\protect\citeauthoryear{Alcabes, Olson  \& Abbot}{Alcabes
  et~al.}{2020}]{alcabes2020robustness}
Alcabes O.~D.,  Olson S.,   Abbot D.~S.,  2020, Monthly Notices of the Royal
  Astronomical Society, 492, 2572

\bibitem[\protect\citeauthoryear{{Alderson} et~al.,}{{Alderson}
  et~al.}{2022}]{aldersonERS2022}
{Alderson} L.,  et~al., 2022, arXiv e-prints, \href
  {https://ui.adsabs.harvard.edu/abs/2022arXiv221110488A} {p. arXiv:2211.10488}

\bibitem[\protect\citeauthoryear{Amundsen et~al.,}{Amundsen
  et~al.}{2016}]{amundsen2016uk}
Amundsen D.~S.,  et~al., 2016, Astronomy \& Astrophysics, 595, A36

\bibitem[\protect\citeauthoryear{Arthur \& Nicholson}{Arthur \&
  Nicholson}{2022}]{gaianhabzone}
Arthur R.,  Nicholson A.,  2022, Monthly Notices of the Royal Astronomical
  Society, in press

\bibitem[\protect\citeauthoryear{Becker \& Sibani}{Becker \&
  Sibani}{2014}]{becker2014evolution}
Becker N.,  Sibani P.,  2014, EPL (Europhysics Letters), 105, 18005

\bibitem[\protect\citeauthoryear{Boutle, Mayne, Drummond, Manners, Goyal,
  Lambert, Acreman  \& Earnshaw}{Boutle et~al.}{2017}]{Boutle:2017}
Boutle I.~A.,  Mayne N.~J.,  Drummond B.,  Manners J.,  Goyal J.,  Lambert
  H.~F.,  Acreman D.~M.,   Earnshaw P.~D.,  2017, Astronomy \& Astrophysics,
  601, 13

\bibitem[\protect\citeauthoryear{Catling, Kiang, Robinson, Rushby, Genio
  et~al.}{Catling et~al.}{2018}]{catlingdavid2018exoplanet}
Catling D.~C.,  Kiang N.~Y.,  Robinson T.~D.,  Rushby A.~J.,  Genio A.~D.,
  et~al., 2018, Astrobiology

\bibitem[\protect\citeauthoryear{Chopra \& Lineweaver}{Chopra \&
  Lineweaver}{2016}]{chopra2016case}
Chopra A.,  Lineweaver C.~H.,  2016, Astrobiology, 16, 7

\bibitem[\protect\citeauthoryear{Christensen, Di~Collobiano, Hall  \&
  Jensen}{Christensen et~al.}{2002}]{christensen2002tangled}
Christensen K.,  Di~Collobiano S.~A.,  Hall M.,   Jensen H.~J.,  2002, Journal
  of theoretical Biology, 216, 73

\bibitem[\protect\citeauthoryear{Claudi.}{Claudi.}{2017}]{Claudi:2017}
Claudi. R.,  2017, Proceedings of Science

\bibitem[\protect\citeauthoryear{Collins}{Collins}{2021}]{collins2021modeling}
Collins M.,  2021, LPI Contributions, 2549, 7001

\bibitem[\protect\citeauthoryear{Daines, Mills  \& Lenton}{Daines
  et~al.}{2017}]{Daines:2017}
Daines S.~J.,  Mills B.~J.,   Lenton T.~M.,  2017, Nature Communications, 8, 1

\bibitem[\protect\citeauthoryear{Downing \& Zvirinsky}{Downing \&
  Zvirinsky}{1999}]{downing1999simulated}
Downing K.,  Zvirinsky P.,  1999, Artificial life, 5, 291

\bibitem[\protect\citeauthoryear{Dyke}{Dyke}{2010}]{dyke2010daisystat}
Dyke J.~G.,  2010, in Artificial Life XII: Twelfth International Conference on
  the Synthesis and Simulation of Living Systems (18/08/10 - 22/08/10). pp
  349--359, \url {https://eprints.soton.ac.uk/272879/}

\bibitem[\protect\citeauthoryear{Fauchez et~al.,}{Fauchez
  et~al.}{2021}]{fauchez2021trappist}
Fauchez T.~J.,  et~al., 2021, The Planetary Science Journal, 2, 106

\bibitem[\protect\citeauthoryear{{Feinstein} et~al.,}{{Feinstein}
  et~al.}{2022}]{feinsteinERS2022}
{Feinstein} A.~D.,  et~al., 2022, arXiv e-prints, \href
  {https://ui.adsabs.harvard.edu/abs/2022arXiv221110493F} {p. arXiv:2211.10493}

\bibitem[\protect\citeauthoryear{Harding}{Harding}{1999}]{harding1999food}
Harding S.~P.,  1999, Tellus B, 51, 815

\bibitem[\protect\citeauthoryear{H{\"o}ning \& Spohn}{H{\"o}ning \&
  Spohn}{2016}]{honing2016continental}
H{\"o}ning D.,  Spohn T.,  2016, Physics of the Earth and Planetary Interiors,
  255, 27

\bibitem[\protect\citeauthoryear{H{\"o}ning, Hansen-Goos, Airo  \&
  Spohn}{H{\"o}ning et~al.}{2014}]{honing2014biotic}
H{\"o}ning D.,  Hansen-Goos H.,  Airo A.,   Spohn T.,  2014, Planetary and
  Space Science, 98, 5

\bibitem[\protect\citeauthoryear{Kasting, Whitmire  \& Reynolds}{Kasting
  et~al.}{1993}]{kasting1993habitable}
Kasting J.~F.,  Whitmire D.~P.,   Reynolds R.~T.,  1993, Icarus, 101, 108

\bibitem[\protect\citeauthoryear{Kharecha, Kasting  \& Siefert}{Kharecha
  et~al.}{2005}]{Kharecha:2005}
Kharecha P.,  Kasting J.,   Siefert J.,  2005, Geobiology, 3, 53

\bibitem[\protect\citeauthoryear{Kiang, Domagal-Goldman, Parenteau, Catling,
  Fujii, Meadow, Schwieterman  \& Walker}{Kiang et~al.}{2018}]{Kiang:2018}
Kiang N.~Y.,  Domagal-Goldman S.,  Parenteau M.~N.,  Catling D.~C.,  Fujii Y.,
  Meadow V.~S.,  Schwieterman E.~W.,   Walker S.~I.,  2018, Astrobiology, 18

\bibitem[\protect\citeauthoryear{Kirchner}{Kirchner}{2002}]{kirchner2002gaia}
Kirchner J.~W.,  2002, Climatic change, 52, 391

\bibitem[\protect\citeauthoryear{Kopp, Kirschvink, Hilburn  \& Nash}{Kopp
  et~al.}{2005}]{kopp2005paleoproterozoic}
Kopp R.~E.,  Kirschvink J.~L.,  Hilburn I.~A.,   Nash C.~Z.,  2005, Proceedings
  of the National Academy of Sciences, 102, 11131

\bibitem[\protect\citeauthoryear{Kopparapu}{Kopparapu}{2013}]{kopparapu2013revised}
Kopparapu R.~K.,  2013, The Astrophysical Journal Letters, 767, L8

\bibitem[\protect\citeauthoryear{Kopparapu, Ramirez, SchottelKotte, Kasting,
  Domagal-Goldman  \& Eymet}{Kopparapu et~al.}{2014}]{kopparapu2014habitable}
Kopparapu R.~K.,  Ramirez R.~M.,  SchottelKotte J.,  Kasting J.~F.,
  Domagal-Goldman S.,   Eymet V.,  2014, The Astrophysical Journal Letters,
  787, L29

\bibitem[\protect\citeauthoryear{Krissansen-Totton, Thompson, Galloway  \&
  Fortney}{Krissansen-Totton et~al.}{2022a}]{Krissansen-Totton:2022}
Krissansen-Totton J.,  Thompson M.,  Galloway M.~L.,   Fortney J.~J.,  2022a,
  arXiv

\bibitem[\protect\citeauthoryear{Krissansen-Totton, Thompson, Galloway  \&
  Fortney}{Krissansen-Totton et~al.}{2022b}]{krissansen2022understanding}
Krissansen-Totton J.,  Thompson M.,  Galloway M.~L.,   Fortney J.~J.,  2022b,
  Nature Astronomy, 6, 189

\bibitem[\protect\citeauthoryear{Lambert \& Kussell}{Lambert \&
  Kussell}{2014}]{lambert2014memory}
Lambert G.,  Kussell E.,  2014, PLoS genetics, 10, e1004556

\bibitem[\protect\citeauthoryear{Lenardic, Crowley, Jellinek  \&
  Weller}{Lenardic et~al.}{2016}]{lenardic2016solar}
Lenardic A.,  Crowley J.,  Jellinek A.,   Weller M.,  2016, Astrobiology, 16,
  551

\bibitem[\protect\citeauthoryear{Lenardic, Weller, H{\"o}ink  \&
  Seales}{Lenardic et~al.}{2019}]{lenardic2019toward}
Lenardic A.,  Weller M.,  H{\"o}ink T.,   Seales J.,  2019, Physics of the
  Earth and Planetary Interiors, 296, 106299

\bibitem[\protect\citeauthoryear{Lenton \& Watson}{Lenton \&
  Watson}{2013}]{lenton2013revolutions}
Lenton T.,  Watson A.,  2013, Revolutions that made the Earth.
OUP Oxford

\bibitem[\protect\citeauthoryear{Lenton, Daines,   \& J.W.}{Lenton
  et~al.}{2018}]{Lenton:2018}
Lenton T.~M.,  Daines S.~J.,    J.W. M.~B.,  2018, Earth-Science Reviews, 178,
  1

\bibitem[\protect\citeauthoryear{Lovelock}{Lovelock}{1965}]{lovelock1965physical}
Lovelock J.~E.,  1965, Nature, 207, 568

\bibitem[\protect\citeauthoryear{Lovelock}{Lovelock}{1990}]{lovelock1990hands}
Lovelock J.~E.,  1990, Nature, 344, 100

\bibitem[\protect\citeauthoryear{Lovelock \& Margulis}{Lovelock \&
  Margulis}{1974}]{lovelock1974atmospheric}
Lovelock J.~E.,  Margulis L.,  1974, Tellus, 26, 2

\bibitem[\protect\citeauthoryear{Nicholson, Wilkinson, Williams  \&
  Lenton}{Nicholson et~al.}{2017}]{nicholson2017multiple}
Nicholson A.~E.,  Wilkinson D.~M.,  Williams H.~T.,   Lenton T.~M.,  2017,
  Journal of Theoretical Biology, 414, 17

\bibitem[\protect\citeauthoryear{Nicholson, Wilkinson, Williams  \&
  Lenton}{Nicholson et~al.}{2018}]{nicholson2018gaian}
Nicholson A.~E.,  Wilkinson D.~M.,  Williams H.~T.,   Lenton T.~M.,  2018,
  Monthly Notices of the Royal Astronomical Society, 477, 727

\bibitem[\protect\citeauthoryear{Nicholson, Daines, Mayne, Eager-Nash, Lenton
  \& Kohary}{Nicholson et~al.}{2022}]{nicholson2022predicting}
Nicholson A.,  Daines S.,  Mayne N.,  Eager-Nash J.,  Lenton T.,   Kohary K.,
  2022, Monthly Notices of the Royal Astronomical Society, 517, 222

\bibitem[\protect\citeauthoryear{{Paradise} \& {Menou}}{{Paradise} \&
  {Menou}}{2017}]{paradise2017}
{Paradise} A.,  {Menou} K.,  2017, \mn@doi [\apj] {10.3847/1538-4357/aa8b1c},
  \href {https://ui.adsabs.harvard.edu/abs/2017ApJ...848...33P} {848, 33}

\bibitem[\protect\citeauthoryear{Pierrehumbert \& Gaidos}{Pierrehumbert \&
  Gaidos}{2011}]{pierrehumbert2011hydrogen}
Pierrehumbert R.,  Gaidos E.,  2011, The Astrophysical Journal Letters, 734,
  L13

\bibitem[\protect\citeauthoryear{Plank \& Manning}{Plank \&
  Manning}{2019}]{plank2019subducting}
Plank T.,  Manning C.~E.,  2019, Nature, 574, 343

\bibitem[\protect\citeauthoryear{Price \& Sowers}{Price \&
  Sowers}{2004}]{price2004temperature}
Price P.~B.,  Sowers T.,  2004, Proceedings of the National Academy of
  Sciences, 101, 4631

\bibitem[\protect\citeauthoryear{Quanz et~al.,}{Quanz
  et~al.}{2021}]{Quanz:2021}
Quanz S.~P.,  et~al., 2021, Experimental Astronomy, pp 1--25

\bibitem[\protect\citeauthoryear{Ramirez \& Kaltenegger}{Ramirez \&
  Kaltenegger}{2016}]{ramirez2016habitable}
Ramirez R.~M.,  Kaltenegger L.,  2016, The Astrophysical Journal, 823, 6

\bibitem[\protect\citeauthoryear{{Rimmer}, {Xu}, {Thompson}, {Gillen},
  {Sutherland}  \& {Queloz}}{{Rimmer} et~al.}{2018}]{rimmer2018}
{Rimmer} P.~B.,  {Xu} J.,  {Thompson} S.~J.,  {Gillen} E.,  {Sutherland} J.~D.,
    {Queloz} D.,  2018, \mn@doi [Science Advances] {10.1126/sciadv.aar3302},
  \href {https://ui.adsabs.harvard.edu/abs/2018SciA....4.3302R} {4, eaar3302}

\bibitem[\protect\citeauthoryear{{Rustamkulov} et~al.,}{{Rustamkulov}
  et~al.}{2022}]{rustamkulovERS2022}
{Rustamkulov} Z.,  et~al., 2022, arXiv e-prints, \href
  {https://ui.adsabs.harvard.edu/abs/2022arXiv221110487R} {p. arXiv:2211.10487}

\bibitem[\protect\citeauthoryear{Schwartzman \& Volk}{Schwartzman \&
  Volk}{1989}]{schwartzman1989biotic}
Schwartzman D.~W.,  Volk T.,  1989, Nature, 340, 457

\bibitem[\protect\citeauthoryear{Schwieterman et~al.,}{Schwieterman
  et~al.}{2018}]{Schwieterman:2018}
Schwieterman E.~W.,  et~al., 2018, Astrobiology, 18, 663

\bibitem[\protect\citeauthoryear{Seager}{Seager}{2013a}]{seager2013exoplanet}
Seager S.,  2013a, Science, 340, 577

\bibitem[\protect\citeauthoryear{Seager}{Seager}{2013b}]{Seager:2013a}
Seager S.,  2013b, Science, 340, 577

\bibitem[\protect\citeauthoryear{{Seager}}{{Seager}}{2014}]{seager2014}
{Seager} S.,  2014, \mn@doi [Proceedings of the National Academy of Science]
  {10.1073/pnas.1304213111}, \href
  {https://ui.adsabs.harvard.edu/abs/2014PNAS..11112634S} {111, 12634}

\bibitem[\protect\citeauthoryear{Seager, Bains  \& Hu}{Seager
  et~al.}{2013}]{seager2013biomass}
Seager S.,  Bains W.,   Hu R.,  2013, The Astrophysical Journal, 775, 104

\bibitem[\protect\citeauthoryear{Snellen et~al.,}{Snellen
  et~al.}{2021}]{Snellen:2021}
Snellen I. A.~G.,  et~al., 2021, Experimental Astronomy

\bibitem[\protect\citeauthoryear{Sousa-Silva, Seager, Ranjan, Petkowski, Zhan,
  Hu  \& Bains}{Sousa-Silva et~al.}{2020}]{sousa_silva_2018}
Sousa-Silva C.,  Seager S.,  Ranjan S.,  Petkowski J.~J.,  Zhan Z.,  Hu R.,
  Bains W.,  2020, \mn@doi [Astrobiology] {10.1089/ast.2018.1954}, 20, 235

\bibitem[\protect\citeauthoryear{{Tsai} et~al.,}{{Tsai}
  et~al.}{2022}]{tsaiERS2022}
{Tsai} S.-M.,  et~al., 2022, arXiv e-prints, \href
  {https://ui.adsabs.harvard.edu/abs/2022arXiv221110490T} {p. arXiv:2211.10490}

\bibitem[\protect\citeauthoryear{Walker et~al.,}{Walker
  et~al.}{2018}]{walker2018exoplanet}
Walker S.~I.,  et~al., 2018, Astrobiology, 18, 779

\bibitem[\protect\citeauthoryear{Ward}{Ward}{2009}]{ward2009medea}
Ward P.,  2009, in , The Medea Hypothesis.
Princeton University Press

\bibitem[\protect\citeauthoryear{Watson \& Lovelock}{Watson \&
  Lovelock}{1983}]{watson1983biological}
Watson A.~J.,  Lovelock J.~E.,  1983, Tellus B: Chemical and Physical
  Meteorology, 35, 284

\bibitem[\protect\citeauthoryear{Weller \& Lenardic}{Weller \&
  Lenardic}{2018}]{weller2018evolution}
Weller M.~B.,  Lenardic A.,  2018, Geoscience Frontiers, 9, 91

\bibitem[\protect\citeauthoryear{Williams \& Lenton}{Williams \&
  Lenton}{2007}]{williams2007flask}
Williams H.~T.,  Lenton T.~M.,  2007, Oikos, 116, 1087

\bibitem[\protect\citeauthoryear{Wood, Ackland, Dyke, Williams  \& Lenton}{Wood
  et~al.}{2008}]{wood2008daisyworld}
Wood A.~J.,  Ackland G.~J.,  Dyke J.~G.,  Williams H.~T.,   Lenton T.~M.,
  2008, Reviews of Geophysics, 46

\bibitem[\protect\citeauthoryear{Worden}{Worden}{2010}]{worden2010notes}
Worden L.,  2010, Ecological Economics, 69, 762

\bibitem[\protect\citeauthoryear{Zakem, Polz  \& Follows}{Zakem
  et~al.}{2020}]{zakem:2020}
Zakem E.~J.,  Polz M.~F.,   Follows M.~J.,  2020, Nature communications, 11, 1

\makeatother
\end{thebibliography}

\end{document}